# A Systematic Mapping Study Addressing the Reliability of Mobile Applications: The Need to Move Beyond Testing Reliability

Chathrie Wimalasooriya✉, Sherlock A. Licorish, Daniel Alencar da Costa, Stephen G. MacDonell

*Department of Information Science, University of Otago, Dunedin, New Zealand*
*emails: chathrie.wimalasooriya@postgrad.otago.ac.nz,*
*{sherlock.licorish, danielcalencar, stephen.macdonell}@otago.ac.nz*

**Abstract**

*Intense competition in the mobile apps market means it is important to maintain high levels of app reliability to avoid losing users. Yet despite its importance, app reliability is underexplored in the research literature. To address this need, we identify, analyse, and classify the state-of-the-art in the field of mobile apps' reliability through a systematic mapping study. From the results of such a study, researchers in the field can identify pressing research gaps, and developers can gain knowledge about existing solutions, to potentially leverage them in practice. We found 87 relevant papers which were then analysed and classified based on their research focus, research type, contribution, research method, study settings, data, quality attributes and metrics used. Results indicate that there is a lack of research on understanding reliability with regard to context-awareness, self-healing, ageing and rejuvenation, and runtime event handling. These aspects have rarely been studied, or if studied, there is limited evaluation. We also identified several other research gaps including the need to conduct more research in real-world industrial projects. Furthermore, little attention has been paid towards quality standards while conducting research. Outcomes here show numerous opportunities for greater research depth and breadth on mobile app reliability.*

**Keywords:** Mapping study, Software reliability, Mobile app reliability, Evidence-based software engineering

## 1. INTRODUCTION

New mobile applications are being developed and released continuously via app stores since the market for mobile devices is both growing and diversifying. Recent statistics show that the Google Play store is home to more than 3.14 million[1] apps and the Apple App store comprises more than 3.4 million apps,[2] up from 2.1 million and 3.0 million just one year ago. The success of any mobile application depends on various quality attributes (QAs), as for any other software system. Careful and constant management of those quality attributes is not only a technical necessity, but also central to a company's survival in the competitive app market [1–3]. In particular, reliability is generally accepted as a key quality attribute since it measures failures and 'misbehaviours' of a software application. Users depend on apps for a wide variety of purposes, from providing entertainment to enabling more serious activities such as m-health, m-commerce and m-government, where greater reliability, availability, responsiveness and performance is essential [4–6] and is expected. Failure of mobile applications in these contexts could make users frustrated, abandon apps and move to alternative apps [7]. This may affect operational revenues/costs and the reputation of the company [8]. As an example, Amazon's app continued to crash on its annual prime day in 2018,[3] where the company recorded significant losses. According to Amazon, a one second delay in responsiveness could cost US$1.6 billion in sales each year [9]. Therefore, paying attention to the reliability of apps, especially when they are in operation after release, is vital to prevent such consequences. As Musa and Everett [10] noted, ''Reliability is probably the most important of the characteristics inherent in the concept Software Quality''. Due to the fact that there is constant growth in the app market, and this domain is becoming more and more competitive [11–13], paying attention to reliability is becoming more important [14–16].

In this work we refer to applications that are not supported to run on the current generation of mobile devices, such as smartphones and tablet PCs, as 'traditional software'. Thus, web or desktop applications (e.g., Eclipse, Apache HTTP, Gitlab) are traditional software. Compared to traditional software, mobile applications are different as they must deal with particular constraints. For instance, mobile applications must be able to run on limited power supply, which demands consideration when assessing the energy efficiency and the performance of mobile hardware [17]. Also, mobile platforms (e.g., Android OS, iOS[4]) are upgraded regularly in relatively short time periods, which increases compatibility issues, possibly leading to app crashes [18]. Due to the fact that mobile apps differ from traditional software, traditional approaches to development and deployment, such as the techniques used for exception handling, testing and automated program repair, cannot be applied without first verifying their utility in a mobile context [19,20]. Hence, over the past few years, much research has been published investigating the reliability of mobile app software (e.g., [1,18,20–22]).

---

[1] https://www.statista.com/statistics/289418/number-of-available-apps-in-the-google-play-store-quarter/.

[2] https://www.statista.com/statistics/268251/number-of-apps-in-the-itunes- app-store-since-2008/.
[3] https://www.theatlantic.com/technology/archive/2018/07/prime-day-amazon-website/565412/.

This body of research essentially addresses two different concepts of reliability. The first concept, operational reliability, is defined as the probability of no failure occurring during the operation of an app [23]. The second concept, testing reliability, is the probability of no failure occurring during the testing phase of the software development process [23]. A study published in 2016 by Zein et al. [20] analysed the literature to that date on mobile app testing approaches that focussed on testing reliability. To the best of our knowledge, no previous work has sought to survey evidence in the area of operational reliability of mobile applications. Hence, it is not clear to what extent the research community has achieved success in terms of operational reliability. This motivated us to conduct a study surveying and synthesizing existing literature. To further motivate this study, we conducted an informal tertiary review to search for systematic secondary studies published on operational reliability of mobile apps, where results showed no such studies were published (refer to Section 3). Such a systematic secondary study would help researchers and practitioners to focus their efforts in achieving better reliability in mobile apps.

To fulfil this need, we conducted a systematic mapping study (SMS) in the field of operational reliability focusing on Android mobile applications. Our focus is on Android apps because Android holds 72.9% of the global market share[5] and Android apps are particularly popular in the domain of mobile apps research, due to Android's open-source software (OSS) nature. Our goal of this study is to collect evidence of all the related literature, structure the evidence according to a classification scheme and identify gaps in the current research and practice landscape that may need further attention. To achieve our goal, we consider three main research questions (RQs). RQ1 explores publication trends (with authors and venues of publication) of the relevant studies, RQ2 explores the nature of the studies based on research types, contributions, quality attributes and metrics used, and RQ3 investigates empirical studies in more detail, considering their research method, study settings, datasets, and finally, the limitations and research gaps in the area.

The contributions of our study are as follows:

- We provide a comprehensive understanding of the state-of-the-art of research on a broad topic, i.e.,

Table 1. SQuaRE reliability attributes definitions.

| Reliability | Degree to which a system, product or component performs specified functions under specified conditions for a specified period of time |
|---|---|
| • Availability | Degree to which a system, product or component is operational and accessible when required for use |
| • Maturity | Degree to which a system, product or component meets needs for reliability under normal operation |
| • Fault-tolerance | Degree to which a system, product or component operates as intended despite the presence of hardware or software faults |
| • Recoverability | Degree to which, in the event of an interruption or failure, a product or system can recover data directly affected and re-establish the desired state of the system |

operational reliability of Android apps. We classified existing work by adapting multiple classification schemes (e.g., classifications based on research focus, contribution and the like). Researchers could use, adapt, or further refine these classification schemes to categorize and describe new research in related domains.

- We provide evidence of existing work in the field and quantify the frequency of these works based on several adapted classification schemes. Quantifying existing works provides an understanding of the extent to which a particular area of focus has been investigated.

- We identify a list of research gaps (e.g., areas that lack evaluation or tool support) in order to outline current research opportunities and suggest areas for further research.

- Practitioners may also use the results of this study to identify existing approaches to handling reliability-related issues as well as to analyse the maturity level and potential risks of various approaches before applying them.

The remaining sections of this paper are organized as follows. Section 2 summarizes the background and insights of previous related studies. Section 3 describes our informal tertiary review. Section 4 describes the methodology we followed, including our research questions and the classification schemes used for data extraction. Section 5 presents the results and identifies gaps in the space, providing recommendations for future research. Section 6 discusses the results and outlines their implications. Section 7 evaluates threats to validity of our study. Finally, Section 8 concludes the work.

## 2. BACKGROUND AND RELATED STUDIES

This section provides background information for understanding the concepts presented throughout this study (in Sections 2.1– 2.3). We then review related works in the area (in Section 2.4).

---

[4] https://developer.apple.com/documentation/ios-ipados-release-notes.

[5] https://www.statista.com/statistics/272698/global-market-share-held-by-mobile-operating-systems-since-2009/.

## 2.1. Software reliability

Research on software reliability has its genesis in the early 1970s [24]. Software reliability is defined as the probability of failure-free software operation for a specified period of time in a specified environment [25]. ISO/IEC SQuaRE [26] provides a broader definition; ''the degree to which a system, product or component performs specified functions under specified conditions for a specified period of time''. In the ISO/IEC SQuaRE context, reliability is considered a combination of four sub-attributes: availability, maturity, fault-tolerance and recoverability. Table 1 provides the standard definitions for these attributes.

As mentioned in Section 1, reliability may be partitioned according to two concepts: testing reliability and operational reliability. During the testing phase, software is improved by removing identified faults, hence, reducing the potential for failures to reoccur. As in O'connor [27], we refer to failures as the unexpected departure from how software should behave during operation according to the requirements. A fault (a property of a program) is a defect in a program that causes failures (a property of the program's execution) when the program is executed. In other words, failure is dynamic, and a program needs to be executed for a failure to occur. Not all faults are detectable by testing approaches. Faults that escape the testing phase (field faults) can cause failures during later operation of the software. Field faults may or may not trigger field failures, depending on the execution context of the production environment. During the operational phase, unlike in the testing phase, it is not the case of removing faults, but users can experience failures (known as field failures). Such failures make operational reliability much more important than testing reliability from users' point of view [23].

Overall, there are four types of techniques (Fault prevention, Fault removal, Fault tolerance and Fault/failures forecasting) that may be applied by software engineers to achieve reliability [25]. (1) Fault prevention: the first mechanism to defend against unreliability is to avoid faults and failures. In general, standard development methodologies, best development practices, programming principles and interacting with users early to verify/refine requirements are the approaches recommended to prevent faults. (2) Fault removal: fault prevention cannot guarantee avoiding all software faults [25]. Once faults are introduced in the software, removing them is the next defencive mechanism. Standard industry practices for removing faults are software inspection [28] and testing [20]. (3) Fault tolerance: faults can slip through all testing and inspection procedures and stay with the software when it is released. In such cases, the last defencive mechanism is fault tolerance, to prevent triggering faults as failures. Usual fault tolerance techniques are: adding exception handling mechanisms to stop further faults propagating, adding conditional checks (e.g., checks for inputs/outputs or illegal operations), and so on. (4) Fault/failure forecasting: finally, if failures occur in the field, it is important to estimate or predict them to understand and assess the quality of the system or when to stop testing. Various models (e.g., Software Reliability Growth models) have been proposed to estimate and predict reliability [29].

Measuring software reliability is also important to plan maintenance activities. For instance, maintenance involving software rejuvenation techniques can be scheduled accordingly [25]. Rejuvenation of software involves stopping the running software occasionally, cleaning the internal state and its execution environment, and restarting it. Removing accumulated errors, freeing up operating system resources and reinitializing internal data structures are some actions involved in cleaning the internal state and execution environment of software. Software rejuvenation is a fault tolerant technique intended to prevent failures caused by degradation due to software ageing (a phenomenon of software performance degradation, leading to slow UI responses, failure-prone states or even app crashes) [30].

## 2.2. Android

**Android OS:** Android[6] is an open-source, Linux-based software stack designed for various types of devices such as TVs, mobile phones, watches and cars. The Android platform is roughly divided into four layers according to its architecture: (1) Linux kernel is the bottom layer, which the entire Android platform is built on. (2) Hardware Abstraction Layer (HAL) is on top of the Linux kernel and has several libraries that implement interfaces exposing device capabilities to the higher-level Java API framework. When the API framework makes a call to access the hardware, the Android system loads libraries relevant to the requested hardware component (e.g., camera, bluetooth). (3) Android Runtime (ART) and native C/C++ libraries are available in the third layer from the bottom. ART also includes core runtime libraries that provide many of the functionalities of the Java programming language. Many Android components such as ART and HAL use native code that require native libraries written in C and C++. (4) Java API framework is a set of APIs written in the Java language that provide a set of higher-level services to applications/app developers in the form of java classes and methods. Through these classes and methods, app developers can access and make use of the features of the Android system. Therefore, research focusing on the Java language can be useful for supporting understandings of Android apps.

**Android apps:** Apps are at the top layer on the Android platform. Apps consist of both system apps (core apps that come with Android for email, calendars, SMS messaging, contacts, internet browsing and more) and developer apps (third party apps). Mobile phone users can install these third-party apps to replace default system apps. ***Distribution and installation of apps:*** For distribution and installation of Android apps, Android uses the format of Android Application Package (APK) files (with the .apk file extension). The .apk file contains code, resources, certificates, assets and AndroidManifest.xml (global configuration file for the app) [31]. ***Manifestation:*** Every .apk file must have an Android-manifest.xml file which contains essential configuration details for each Android app. The manifest file must declare the unique application ID, permission details that an app needs to access protected system properties or other apps and hardware/software details (e.g., version of Android devices on which apps can be installed) that the app requires [32]. Manifest files also include properties of four basic components of Android apps (Activities, Services, Broadcast Receivers and Content Providers) that are required to control execution behaviour of the components. An app developer can define how to launch the components through Intent which is a messaging object that facilitates communication between different components [32]. *App programming model:* Android apps are primarily built around the concept of activities. An activity represents a single screen/page with UI components and logic behind the screen. When users interact with an app, an activity goes through a sequence of states in the activity lifecycle, such as 'created', 'started', 'resumed', 'paused', 'stopped' and 'destroyed'. App developers can customize app transition through these lifecycle states by overriding the lifecycle callback methods provided by the Android system [33,34].

---

[6] https://developer.android.com/guide/platform.

**2.3. Reliability of Android apps**

In addition to the traditional techniques (e.g., exception handling, rejuvenation, see Section 2.1), Android apps also require more Android-specific techniques to ensure app reliability [18, 20]. For example, there can be failures due to improper permission definitions in the manifest file. Developers may make mistakes while defining permissions to system resources, which may lead to app crashes while apps are running. Furthermore, app developers must configure the device parameters (e.g., minSdkVersion, maxSdkVersion) in the manifest file which indicates the supported API-level as recommended in Android documentation.[7] This practice is not yet established in the app development community, however, not setting such parameters or setting wrong versions can lead to compatibility issues including crashes. Moreover, Android apps use event-driven programming where the flow of the program is determined by a user's actions (mouse clicks, key presses). Modern GUI frameworks such as Android, SWT and Swing employ a single-thread-model to process user events [35]. When an application is launched, the model creates a UI thread, in which it will run the application. To maintain a responsive user interface, it is important not to block the UI thread. The common practice to do so is to offload intensive tasks to background threads. This means making asynchronous threads (i.e., background threads) to perform long-running tasks such as computationally or resource-intensive tasks (e.g., network access, database queries). Android Development Framework (ADF) provides several asynchronous programming constructs (e.g., AysncTask, Thread, IntentService) to achieve this goal. However, developers can still use inappropriate asynchronous programming constructs which may in turn cause operational errors [35].

**2.4. Related work**

We identified several surveys and systematic literature reviews (SLRs) related to software reliability [29,36–39]. However, these studies investigate traditional software, with none specifically reviewing the state-of-the-art of reliability in mobile applications. For instance, Singhal and Singhal [36] performed an SLR study that identified the state-of-the-art research in the field of software reliability, covering literature published up to 2011. This was a decade ago, when mobile applications were still becoming mainstream. The study classifies 141 papers according to research topic, research approach (e.g., survey, theory) and study context (e.g., academic, industry). Their findings suggested that more industrial research was required since existing evidence at the time was not sufficient to show industrial validity. With respect to methodology the authors highlighted the importance of manual search to find relevant literature in the field, due to a lack of a standardized use of terms pertaining to software reliability.

Shahrokni and Feldt [37] conducted a literature review in the field of software robustness — which is defined as a reliability characteristic in some standards; for instance, IEEE-STD 610.12-1990-covering the period from 1990–2010. The authors analysed and categorized 144 papers based on development phase (e.g., requirements, design and implementation), domain (e.g., web application, distributed application), research type (e.g., evaluation, experience report), contribution (e.g., tool, metrics), and evaluation type (e.g., academic, industrial). The main gap identified in the study was the lack of research on elicitation and specification of software robustness requirements. Another finding was that most of the studies considered only one aspect of robustness (invalid inputs), ignoring all other complex aspects such as unexpected events, timeout, interrupts and stressful execution environments.

Febrero et al. [29] conducted a SMS of software reliability modelling and analysed 503 studies covering the period from 2003 to 2014. They grouped reliability models into five classes: Software Reliability Growth models, Bayesian methods, Test-based methods, Artificial Intelligence-based techniques, and Static and Architectural Reliability models. The study found a gap in research where many studies were not conducted following the established quality standards. To fill this gap, the same authors conducted an SLR [39] on software reliability assessment based on quality standard ISO/IEC −25000 SQuaRE, covering the period 1991–2014. That latter study showed that consideration of quality standards and reliability from different perspectives to meet different stakeholders' needs received very little attention. They also mentioned that the main drawback of the extant reliability models was that they were too complex to apply in daily practice. Another challenge in developing effective models was the lack of consensus and a wide variety of views on what reliability means. For instance, the authors observed that the terms 'reliability' and 'dependability' were often used interchangeably and even as synonyms. While our work studies the state-of-the-art of reliability in mobile apps, their focus was on how reliability models applied reliability standards (e.g., ISO/IEC-25000 SQuaRE). In addition, our study surveys more recent work in the field (six or more years later).

A more recent SMS conducted by Alhazzaa and Andrews [38] focused on reliability growth models that consider the evolution of software systems. They summarized trends with respect to year of publication, venues, and study context (academic, industry). Studies were classified according to solution extent (the type and quantity of change: single change-point, multiple change-points), proposed method (analytical and curve-fit) and research type (empirical or non-empirical). Further, they evaluated the quality of empirical studies according to the evaluation criteria proposed by Ali et al. [40]. They recommended that researchers should aim to provide better quality empirical studies with greater industrial involvement. Also, these authors suggested that future works should investigate the following questions: how long can these models predict into the future? And when do practitioners have to use a different model or update its parameters? All of these previous studies (including the study by Alhazzaa and Andrews [38]) have noted that solutions were evaluated mostly in academic environments without involving or collaborating with practitioners during their research, and thus, lack industrial level validation.

---
[7] https://developer.android.com/training/basics/supporting-devices/platforms.

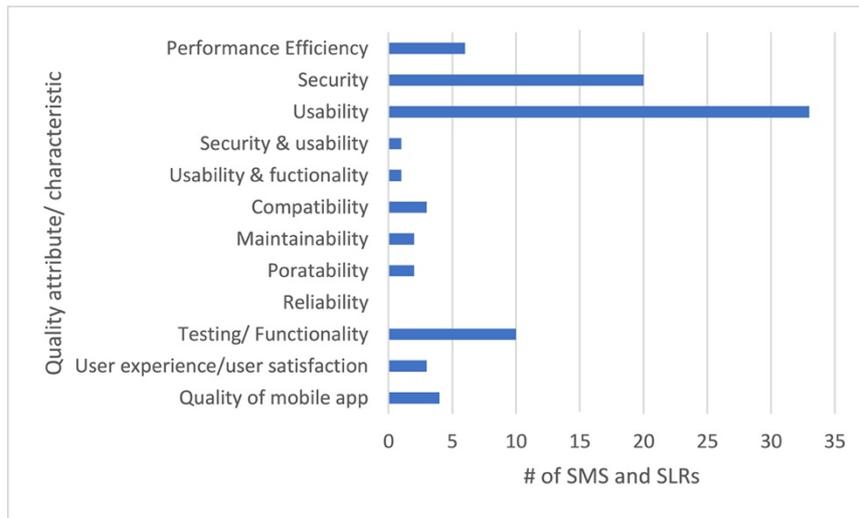

Figure 1. Number of SMSs and SLRs on QAs of mobile applications.

## 3. INFORMAL TERTIARY REVIEW

None of the studies mentioned in the previous section (Section 2.4) have been conducted with a focus on the reliability of mobile applications. Also, we found very little evidence related to reliability of mobile apps compared to other QA (Quality Attribute) reviews that were conducted on mobile apps. This leads us to explore how QAs have been considered in mobile apps-related research. We conducted an informal tertiary review to analyse what QA standards were considered in mobile app review studies. We considered it important to understand whether quality-related research of mobile apps ignore QAs defined in quality standards, since standards play an important role in achieving real world goals [39]. We refer to QAs in this study as the attributes defined in standard quality models such as ISO/IEC 25010:2011 and ISO/IEC 9126. We also use just the term ''attribute'' to represent quality attribute.

To conduct the tertiary review, we used several search terms to cover a wide scope. These search terms can be split into three sets, representing: domain (''mobile'', ''app'', ''Android'', ''iOS''), the type of study (''systematic literature review'', ''mapping study'', ''state-of-the-research'', ''state-of-the-art'', ''state of the research'', ''state of the art'') and context (''quality'', ''non-functional requirement'', ''non functional requirement''). To represent context, we also used the quality dimensions defined in standards ISO/IEC 25010:2011 or ISO/IEC 9126: ''functionality'', ''performance efficiency'', ''usability'', ''reliability'', ''portability'', ''maintainability'', ''performance'', and ''compatibility''. We then further supplemented these terms with other quality-related terms: ''bug'', ''defect'', ''crash'', ''test'', ''anti-pattern'', ''smell'', ''vulnerability'', ''security'', ''energy'', ''user experience'' and ''user satisfaction''. Using these search terms, we performed an automated search on four databases: IEEE Xplore, Springer, Scopus and ACM DL for the ten years 2010–2020. From the automatic search results, we manually selected SMSs and SLRs related to mobile app quality based on paper titles and abstracts. Thereafter, we map the QAs that were the focus of the papers to QAs in the ISO/IEC 25010:2011 standards and classified them. Fig. 1 represents the results of our tertiary review, showing the number of review papers (SMSs and SLRs) found against each quality attribute.

As Fig. 1 shows, our tertiary review discovered a research gap in that there are no previous systematic secondary studies related to the reliability of mobile apps. Therefore, we were interested to see how reliability has been investigated by researchers in a mobile context, what metrics are used, what reliability attributes (i.e., sub-QAs of reliability as defined in standards, such as 'availability' and 'fault-tolerance') are studied and what areas need further research. This knowledge is important to the research and practitioner community who deal with the reliability of mobile apps, given their pervasive use. This way, we may derive consensus on how reliability can be assessed and identify research gaps that will further the discipline. The quality standards provide limited information about how to deal with QAs. Hence, synthesized knowledge based on broad analyses covering state-of-the-art research is needed to understand and support the body of knowledge on reliability. This motivated us to conduct this mapping study to understand the state-of-the-art of research in this field.

# 4. METHODOLOGY

As noted by Kitchenham and Charters [41], a systematic mapping study (SMS) is also referred to as a ''scoping study'', which is a form of systematic literature review (SLR). An SLR is a ''methodologically rigorous review of research results'' [41]. Mapping studies aim to report a broader overview of published studies related to a particular field, and so they do not analyse primary articles in as much depth as in SLRs. Furthermore, mapping studies identify, cluster and report evidence in a domain at a higher-level of granularity to direct the focus of future SLRs and primary studies. Thus, given the lack of previous reviews on the topic of the reliability of mobile apps, an SMS was most appropriate for our investigation.

This section presents the systematic mapping method and protocol that was applied to conduct this study. The protocol was designed and developed during several discussions among the authors by following the guidelines provided by Petersen et al. [42] and Kitchenham and Charters [41]. The protocol includes nine major steps: (1) defining research questions, (2) perform manual search to generate a test set, (3) pilot selection criteria and search terms, (4) construct automatic search, (5) formalize protocol, (6) conduct automatic search, (7) select studies to be included, (8) snowballing and (9) data extraction. We now detail how these steps were implemented in this study. All Appendices (A to E) and data related to our mapping study are available in our online repository [43]. We refer to each specific Appendix when we explain our methodology and results later in this study.

## 4.1. Research questions

To achieve the main goal of this study, which is to understand the state-of-the-art in the field of mobile apps reliability, we derived three main research questions. As mentioned in Section 1, by reliability, we mean ''operational reliability''. Through the first research question (RQ1) we aim to analyse trends of research interest in terms of venues and authors over time. The next two RQs (RQ2 and RQ3) are aimed at building a classification scheme through the identification of related studies, structuring knowledge gained from these studies, and also to identify research gaps that could extend the current body of knowledge. We define the research questions as follows.

**RQ1: When and by whom has reliability of Android apps been studied?**

  RQ1.1: Who are the most active authors and which venues publish work on this topic?
  RQ1.2: How has interest in the reliability of Android apps field changed over the years?

*Rationale:* One of the conventions in presenting results of a mapping study is to summarize the authors, venues, and time (published year) of literature to analyse any trends in research interest [42,44]. Answering this RQ provides an indication for researchers about which venues and authors are likely to be involved in similar research, directing networking possibilities, future publication targets and meta-analyses. Also, analysing trends of publishing years illustrate how active the topic has been historically and its current interest.

**RQ2: How has the reliability of Android apps been studied?**
RQ2.1: What is the research focus?
RQ2.2: What is the research type?
RQ2.3: What are the research contributions?
RQ2.4: What reliability attributes have been explored?
Q2.5: What metrics/measures are used to assess reliability?

*Rationale:* The investigations in this RQ help us to build a classification scheme (which we explain in Section 4.10) that enables us to divide and group existing evidence of related research and knowledge gained from the literature. Classifying literature is one of the goals of mapping studies, especially where such evidence has not been organized previously (see Section 2). Furthermore, considering the Quality Standard ISO-25010,8 we examine what reliability-related quality attributes (reliability attributes) are studied, since it is important to know to what extent standard quality models are considered by the research community (see Section 4.10.4). Finally, we look into what metrics and measures have been used to achieve the research objectives of these studies (see Section 5.2.5). These metrics help to realize how reliability can be assessed or quantified, potentially discerning gaps in the domain of measures or specific strengths that may be transferable to industry in their assessment of the reliability of mobile apps.

**RQ3: Which studies have investigated the reliability of Android apps empirically?**
RQ3.1: What research methods do these empirical studies use?
RQ3.2: In what settings are these empirical studies conducted?
RQ3.3: What sources of information/data are used to conduct the empirical investigations?
RQ3.4: What limitations and gaps exist in the empirical studies that could inform future research?

*Rationale:* This RQ is designed to provide an understanding of how reliability has been empirically investigated. Only empirical studies are analysed to answer this RQ, given the drive to gather the aforementioned evidence. The work here analyses what research methods, study environments (academic, industrial), and data have been used in these studies. Such evidence will also inform the last sub-question that explores limitations of current empirical research. We provide directions for further research, but also map out the previous landscape of methods, settings and data for reliability of Android apps research. This is especially novel for the space given the lack of previous similar work (refer to Section 2).

## 4.2. Initial manual search

We developed and piloted a study protocol to conduct our mapping study as suggested in the guidelines. We needed a test set as a reference point to verify our results during piloting the study protocol. To build a test set, we manually searched for relevant papers against a selected list of top ranked venues in the field of software engineering and software reliability for the last 6 years (2015–2021). The selected venues (14 venues) for the manual search are ten conferences: ICSE, ICSME, ASE, MSR, ESEM, FSE, ISSRE, QRS, PRDC, RAMS; and four journals: Empirical Software Engineering, Journal of Systems and Software, IEEE Transactions on Reliability and IEEE Transactions on Software Engineering. We manually searched the dblp websites of these venues and filtered papers based on the titles in the first round. If we were unsure about any paper, we selected it in the first round, resulting in a set of 161 papers. In the second round we went through the full text of the papers to see whether the papers were relevant and would be helpful for answering our RQs. This search was conducted by two authors independently and results were merged later. Inter-rater agreement was calculated using Cohen's Kappa [45], which was found to be 0.65, indicating a good agreement between the two authors based on the categories provided by Landis and Koch [46]. We removed 122 irrelevant records in the second round (these records were marked to be excluded by both authors) and disagreements were left to the final round. The final test set included 39 studies that were selected after discussing and reaching consensus between the two authors. The test set is available in our online repository (Appendix A), included for replication purposes [43].

**4.3. Search string and search strategy**

To conduct the automatic search, we constructed the search string as follows using the steps suggested by Kitchenham and Charters [41].

- Derive major search terms based on the studies from manual analysis (Section 4.2)
- Derive synonyms and alternative terms that expand major search to cover a large area
- Add synonyms and alternatives to each major term with Boolean OR
- Construct the search string by connecting major terms together with Boolean AND

The search string is based on two generic search terms we detected when perusing the 39 studies returned in Section 4.2, ''mobile application'' and ''reliability''. These terms are used to produce relevant results to achieve our research goal. We then expanded the search terms by adding alternative terms and synonyms that we identified from the manual search and the test set. We went through the abstracts of the papers from the test set and noticed that ''crash'' is a popular term when it comes to operational reliability. We thus expanded the term ''reliability'' by adding the term ''crash'' and its stemmed versions. The second major term ''mobile application'' was expanded by incorporating the terms ''smartphone'', ''mobile software'', ''app'' and ''Android'' (since our focus is on Android applications).

We piloted the search strings against the known papers in the test set until the majority of relevant papers were returned. The final search string chosen is the one that returned most of the papers in the test set. Below is the final search string chosen which returned 34 out of 39 papers from the test set, i.e., 87% of our test set.

```
title | abstract | keywords =
((((''Android'') AND (''mobile app'' OR
''mobile apps'' OR ''smartphone app'' OR
''smartphone apps'' OR ''mobile
application'' OR
''mobile applications'' OR ''smartphone
application'' OR ''smartphone
applications'' OR ''mobile software''))
OR
(''Android application'' OR ''Android
applications'' OR ''Android app'' OR
''Android apps''))
AND (''reliability'' OR ''crash*'')
```

Based on the test set, we noticed that although the terms ''availability'', ''maturity'', ''fault-tolerance'' and ''recoverability'' are sub-quality attributes of ''reliability'' (according to the standards ISO/IEC 25010 and ISO 9126), those terms were not used widely in reliability studies. Also those potential search terms contain several other meanings that cause thousands of irrelevant records to be returned by our searches. For example, the terms ''availability'', ''reliability'' or their stemmed versions have a very wide scope and return papers containing ''reliable framework'', ''availability of tools'', ''reliable steps'' or ''reliable method''. Such studies did not consider software reliability, or mobile app reliability. Therefore, we avoided these terms in our search strings, and instead, we manually searched for additional related studies through snowballing (see Section 4.7).

The search string was applied to the databases in Section 4.4 in searches of the title, abstract and keywords of the papers. We restricted our search only to the ''computer science'' discipline depending on the filtering facilities provided by the databases (e.g., Springer Link, Scopus), since the term ''mobile'' is also used in other research disciplines such as mathematics, physics and medicine. Furthermore, the search mechanisms are not consistent across every database [44]. Therefore, we had to adapt the syntax of our searches, and also the search mechanism, depending on the features provided by the databases. For instance, ScienceDirect is not supported with complex search queries, so the search strings were applied in parts for several rounds of searches, and the results were merged subsequently. Another such instance is Springer Link, where filtering based on abstracts and keywords is not supported. Therefore, searches were conducted on the full text. According to Springer settings, searches must be applied either on the 'full text' or 'titles' of the papers. Since searches only on the ''Title'' may be shallow and ineffective (simple search queries can only be applied on filtering based on ''Title'' which returns thousands of irrelevant results) [44], we applied the searches against the full text. This returned 2803 records for the final search, given that if any of our terms was detected in the full text of the study, the paper was added to the results pool.

Note that Springer searches are weighted depending on where the search terms appear in papers, which means that papers with search terms in the title will appear first, then the papers with search terms in the abstracts, then the papers with search terms in the full text (this information was provided by Springer library support team on 2nd May 2020). In order to handle this situation with Springer, we followed strategies inspired by previous studies. For example, Maplesden et al. [44] considered only the first 2000 results out of 26,677 results (7.5% of the returned results) and Savolainen et al. [47] considered only the first 500 results if a search returned more than 1500 results (33% of the returned results). Like these previous studies, we checked only the first 1400 records (50% of the returned 2803 records) records after screening the results in the order of relevance provided by Springer. We believe this strategy is effective since we were unlikely to find a paper of relevance after about the first 703 records. This is because we found only one paper between the 632nd and 703rd record during the process of a manual check of selected 1400 records based on their titles, abstracts and full texts (when necessary) during our study selection process (see Section 4.6). There were no relevant papers even for the rest of the 693 (1400-703) records until the 1400th record. Table 2 summarizes the results of our automatic search process with the number of results returned and exact search string applied on each database.

### 4.4. Search scope

- **Search venues:** We selected five databases that were used in similar work that focussed on systematic literature reviews and mapping studies in software engineering [29,39].
    1. IEEE Xplore
    2. ACM digital library (DL) 3. ScienceDirect
    3. Springer Link
    4. Scopus

    IEEE Xplore, ACM DL, ScienceDirect and Springer have been used in recent secondary studies in the area of mobile applications (e.g., [13,20,48]). We also included Scopus as recommended by Kitchenham [49], since Scopus is a very useful and powerful database in software engineering research for conducting reviews. Scopus also has a higher degree of overlap with other venues such as EI Compendex and Inspec [44]. There are other databases such as Kluwer Online (this has been merged with Springer Link), Wiley Online and ISI Web of Science. The latter two databases are not as popular as the other aforementioned databases for software engineering. Moreover, Scopus is likely to pick up relevant studies in these databases [50,51].

- **Time period:** Most databases (except Springer) returned records only after 2009 or 2010 by default (without setting a time) for the final search string. In Springer also, the results before 2009 were irrelevant. This may be due to the fact that Android was popular only after its first official release in 2008.9 Therefore, we set the year 2008 to be the search start. Our searches cover the period from January 2008 to July 2021.

### 4.5. Selection criteria

The final search string (as defined in Section 4.3) produced 2032 results including many irrelevant records. In order to focus and select only the relevant papers to answer our research questions, we defined the following Inclusion (I) and Exclusion (E) criteria. As suggested by Kitchenham and Charters [41], we used the test set to pilot, refine and enhance the criteria. These criteria were applied in our study selection process in Section 4.6.

The inclusion criteria of our mapping study are as follows:

- I1. The paper is written in English.
- I2. The paper is published in a fully peer reviewed venue.
- I3. The paper is fully accessible.10
- I4. The paper focuses on operational reliability of Android mobile applications.
- I5. The paper is in the domain of software engineering.

We used following exclusion criteria to exclude a study:

- E1. A summary of a conference/workshop, thesis, editorials, keynote.

- E2. Papers that investigate applications that are intended for use on other embedded devices (e.g., smart TV, smart watch) rather than on mobile devices such as smartphones or tablets.

- E3. Papers that investigate reliability focusing solely on lower levels of the Android operating system (OS) (e.g., [52]), processor architecture [53], or communication infrastructure (e.g., mobile cloud computing solutions [54]). Such investigations require specific focus on other specialized areas such as special hardware constraints (e.g., Linux kernel, memory drivers) and runtime virtual environments which are distinct from our research focus.

- E4. Papers that study tools or practices that are only peripherally related to operational reliability. For example, papers that present strategies to enhance testing techniques to handle defects in general (e.g., [55,56]), without paying attention to specific failures such as crashes. In this regard, papers reporting testing approaches that do not focus on operational reliability or reliability attributes (e.g., availability, fault-tolerance) are excluded. For example, we exclude studies that focus on functional testing [57], usability/GUI testing [58] or regression testing [59], but include papers with testing approaches that focus on reliability issues that can occur while apps are in operation (e.g., some context-aware testing approaches that focus on network disruption which may cause an app to become unresponsive).

### 4.6. Selection process

All selection criteria were discussed among the four authors while developing and refining them until no remaining disagreements were recorded. Furthermore, the authors had several consensus meetings (about 12 meetings) and agreed on a concrete selection process which was used to select the included set of papers in this mapping study. Below we describe this selection process and the steps are summarized in Fig. 2.

As in Fig. 2, first, we conducted the automatic search (Section 4.3) on the five selected databases (Section 4.4) by applying the search terms on papers' title, abstract and keywords. In each database, we restricted the search to the computer science field (where possible) and set the start year as 2008. After the automatic search, we manually analysed the results to remove duplicates. During the process of removing duplicates, if one paper was a journal version of a conference paper, we included the journal version and removed its conference version. In the next phase, we filtered papers based on the selection criteria (Section 4.5). This filtering process has three main steps: in step 1, the full set of papers were filtered based on titles and abstracts while applying all selection criteria (I1 to I5 and E1 to E4). When it was difficult to decide whether the paper should be included or not, such papers went to step 2. In step 2, the papers remaining from the first step were filtered based on the introduction, methodology and conclusion by applying the criteria I4, I5, E2, E3 and E4. Papers that still created uncertainty regarding their inclusion were retained to go to step 3. Finally, in step 3, studies that remained after step 2 were filtered based on a complete reading of the full text while applying I3, I4, E2, E3 and E4. If difficulties or confusions arose while filtering papers, the authors together discussed and made a final decision to include or exclude a paper.

Table 2. Search strings applied on each database and relevant studies returned.

| Database | Search string | # of studies |
|---|---|---|
| IEEE | ((("Document Title":"Android" AND ("mobile app" OR "mobile apps" OR "smartphone app" OR "smartphone apps" OR "mobile application" OR "mobile applications" OR "smartphone application" OR "smartphone applications" OR "mobile software")) OR ("Document Title": "android application" OR "android applications" OR "android app" OR "android apps")) OR (("Abstract":"Android" AND ("mobile app" OR "mobile apps" OR "smartphone app" OR "smartphone apps" OR "mobile application" OR "mobile applications" OR "smartphone application" OR "smartphone applications" OR "mobile software")) OR ("Abstract":"android application" OR "android applications" OR "android app" OR "android apps")) OR (("Author Keywords":"Android" AND ("mobile app" OR "mobile apps" OR "smartphone app" OR "smartphone apps" OR "mobile application" OR "mobile applications" OR "smartphone application" OR "smartphone applications" OR "mobile software")) OR ("Author Keywords":"android application" OR "android applications" OR "android app" OR "android apps"))) AND (("Document Title": "reliability" OR "crash*") OR ("Abstract": "reliability" OR "crash*") OR ("Author Keywords": "reliability" OR "crash*")) | 165 |
| Springer | ((("Android") AND ("mobile app" OR "mobile apps" OR "smartphone app" OR "smartphone apps" OR "mobile application" OR "mobile applications" OR "smartphone application" OR "smartphone applications" OR "mobile software")) OR ("Android application" OR "Android applications" OR "Android app" OR "Android apps")) AND ("reliability" OR "crash*") | 1400[a] |
| Scopus | TITLE ABS KEY(((("Android") AND ("mobile app" OR "mobile apps" OR "smartphone app" OR "smartphone apps" OR "mobile application" OR "mobile applications" OR "smartphone application" OR "smartphone applications" OR "mobile software")) OR ("Android application" OR "Android applications" OR "Android app" OR "Android apps")) AND TITLE-ABS-KEY ("reliability" OR "crash*") | 343 |
| ACM | (Title:(((("Android") AND ("mobile app" OR "mobile apps" OR "smartphone app" OR "smartphone apps" OR "mobile application" OR "mobile applications" OR "smartphone application" OR "smartphone applications" OR "mobile software")) OR ("Android application" OR "Android applications" OR "Android app" OR "Android apps")) OR Abstract: (((("Android") AND ("mobile app" OR "mobile apps" OR "smartphone app" OR "smartphone apps" OR "mobile application" OR "mobile applications" OR "smartphone application" OR "smartphone applications" OR "mobile software")) OR ("Android application" OR "Android applications" OR "Android app" OR "Android apps")) OR Keyword: (((("Android") AND ("mobile app" OR "mobile apps" OR "smartphone app" OR "smartphone apps" OR "mobile application" OR "mobile applications" OR "smartphone application" OR "smartphone applications" OR "mobile software")) OR ("Android application" OR "Android applications" OR "Android app" OR "Android apps"))) AND (Title:("reliability" OR "crash*") OR Abstract: ("reliability" OR "crash*") OR Keyword:("reliability" OR "crash*")) | 84 |
| ScienceDirect | Title, abstract, keywords: ((("Android") AND ("mobile app" OR "mobile apps" OR "smartphone app" OR "smartphone apps" OR "mobile application" OR "mobile applications" OR "smartphone application" OR "smartphone applications" OR "mobile software"))OR ("Android application" OR "Android applications" OR "Android app" OR "Android apps")) AND ("reliability" OR "crash*") | 40 |
| Total # of studies from automatic search | | 2032 |
| After removing duplicates | | 1253 |

[a]See discussion on Springer Link in Section 4.3.

### 4.7. Snowballing

In order to complete the list of selected papers from the above process in Section 4.6, we further performed the snowballing process to identify missing papers that are relevant [42]. During snowballing, we checked the reference list (backward snowballing) and citation list of each selected paper (forward snowballing). In this process we followed the same three steps (based on title, abstract and full text) exactly in the same way as described in the selection process in Section 4.6. The papers selected from snowballing (17 papers) were finally combined with the papers from the database selection process.

### 4.8. Search and selection results

Fig. 2 shows the results of our search and selection process. The Appendix provides the full list of selected papers. In total, 2032 papers were returned from the automatic searches on the databases, where 1253 papers remained after removing duplicates and 65 papers remained after the selection process in which the selection criteria were applied. We added 5 papers [s75,s76,s77,s86,s87] that were found from our manual search (see Sections 4.2 and 4.3). These 5 papers were not found during the automatic search as their titles, abstracts and keywords did not include the terms ''Android apps'' (in [s77,s86,s87]), ''crash'' or ''reliability'' (in [s75] and [s76]). Most of these papers were published in the last year (e.g., [s75,s86,s87]) which made it unlikely that they would be found through snowballing ([s77], which was published a couple of years ago, was found by snowballing). Snowballing was useful nonetheless, as we added 17 more papers through this process. Thus, in total, 87 (65+ 17+ 5) [43] papers were finally selected for analysis in this mapping study. See the ''List of selected studies'' at the end of this paper for full details of the selected papers.

### 4.9. Data extraction

Each selected paper was reviewed and classified according to a classification scheme, enabling us to create a structure around the included papers and also to systematically extract relevant answers for our research questions. We extracted the data items listed in Table 3 from each selected study.

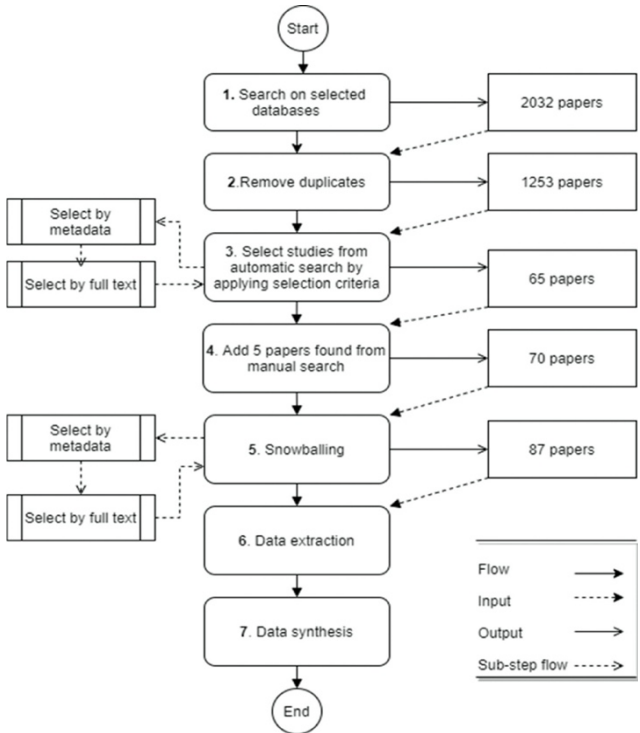

Figure 2. Selection process and total papers returned for synthesis.

Table 3. Data items extracted from each study.

| Internal Reports and Meeting Minutes | Communication |
|---|---|
| Memos of Ministry of Education (MoE) | Correspondence between representatives of MoE and Exec. of the main vendor |
| Cabinet Meeting Minutes | Correspondence emails of end users |
| Status Reports | **Project Related Documents** |
| Steering Committee Meeting Minutes | Risk Registers |
| Payroll Reference Group Meeting Minutes | Project Initiation Doc. |
| Novopay Board Meeting Minutes | Request For Proposal Doc.- including revised versions |
| Quarterly Meeting Reports for High Risk Projects | Business Case Doc. |
| **Periodic Reports of External Companies** | Fallback plans and proposals |
| | Test Plans and strategies |
| PwC | |
| Deloitte | Communication plans |
| Equinox | Reports about variations in the agreement |
| Maven | Progress review reports |
| IQANZ | Surveys carried out from end-users |
| Extrinsic | Remedial plans and programs |

To answer RQ1, we extracted data such as authors' names, affiliations, publication year, publication type and venue. The extraction of this data was straightforward, and thus there was no need for a classification scheme. Data regarding research focus, research approaches, research methods, contribution types, study setting, quality attributes and metrics used were extracted to answer RQ2 and RQ3. Many of these data required the use of classification schemes (refer to Section 4.10). During extraction, the data was stored in a spreadsheet structured to map to each research question. We answered the research questions based on frequencies of papers (frequencies for each category in the classification scheme when necessary). Additionally, the limitations and challenges reported in the papers were recorded and summarized. Furthermore, the overall classification scheme with resulting papers helped us to identify more research gaps and synthesize our findings.

### 4.10. Classification schemes

As inspired by Petersen et al. [42], the keywording process was applied on the selected papers to form our classification scheme. The keywording process has two phases. In the first phase, the main researcher reads the abstracts and keywords of each paper and identifies a set of keywords and concepts that reflect the investigated problem, focus and contributions of the studies. When the abstract is too short or not clear enough for keywords to be chosen, the researcher reads the introduction, methodology and conclusion sections as well. In the second phase, the identified keywords and concepts from different papers are combined (considering alternative keywords and merging similar concepts) to form higher-level concepts that should help readers to understand the context of the papers. These concepts assist the derivation of a set of topics and sub-topics to form categories in the classification scheme representing the underlying research.

In Table 3, we classified papers according to seven different dimensions: (1) research focus (related to RQ2.1), (2) research type (related to RQ2.2), (3) contribution type (RQ2.3), (4) quality attributes (RQ2.4), (5) research method (RQ3.1), (6) study setting (RQ3.2), and (7) details of data considered in investigations (RQ3.3). These categories are now described.

#### 4.10.1. Research focus

The first classification structures the focus area of the selected papers and is used to answer RQ2.1. From the first phase of the keywording process, a large number of concepts were generated due to the diversity of the problems investigated in the papers. For example, crash reproduction, crash detection, test case generation, event handling, run-time change handling, modelling software failures and context-awareness are such concepts. In the second phase, the concepts were grouped together and seven main concept types were derived representing the research focus areas: failure/crash analysis, exception handling, ageing and rejuvenation, API-related issues, self-healing, and runtime change handling. If a study did not fit into any of those categories, it was grouped under the "other" category. The main focus of other category papers is not reliability but was partly related to reliability.

#### 4.10.2. Research type

The second classification is based on the type of "research approach" employed and is used to answer RQ2.2. As inspired by Petersen et al. [60] and Wieringa et al. [61], each selected paper was classified as one of the following:

- **Solution proposal:** papers that propose a novel solution or a significant improvement to an existing

approach that supports maximizing or improving the reliability of mobile apps. These papers usually explain the components of a proposed solution, illustrate how components work using an example, and argue for its applicability without a full validation.

- **Validation research:** papers that propose or investigate a solution (which may be a solution proposed by others) that has not yet been used in practice. Also, it should include a full-blown validation to prove applicability of the solution.

- **Evaluation research:** papers that investigate existing approaches in practice. These papers usually evaluate existing methods (rather than proposing them) by applying them in different contexts, or in this case they might evaluate mobile apps' reliability by using existing methods or tools in practice.

- Philosophical research: papers that look at existing approaches in a new way by structuring the field as a new conceptual framework.

- **Experience report:** papers that explain the personal experiences of the author(s). These papers usually contain a list of lessons learned from using a tool or technique in practice. The experience is reported without a discussion of research methods.

- **Opinion paper:** papers explaining personal opinions of the author(s) on a particular technique, whether it is good or bad, how things could have been done or what should or should not be done. These studies do not rely on related work and research methodologies.

- **Secondary study:** this category is added by us to cover secondary studies that review existing research.

As discussed by Petersen et al. [60], we use solution proposals, philosophical research, and opinion papers to classify non-empirical research.

### 4.10.3. Research contribution

Our classification for contributions, which is used to answer RQ2.3, is inspired in part by Shahrokni and Feldt [37]. Those authors studied papers related to ''robustness'', which is similar to our topic of interest (i.e., reliability). Their work also helps us to distinguish between robustness and reliability. The categories used by Shahrokni and Feldt [37] are: framework, method, tool, metrics, model, evaluation and review. In addition to the contribution types used by Shahrokni and Feldt [37], we included two more types: (benchmark and taxonomy). Below we briefly explain each contribution type.

- **Framework:** a conceptual structure that provides guidance for a detailed method that covers a wide area (e.g., answering several research questions). A framework may be partially automated.

- **Method:** a process (or technique/method) that has a specific goal that covers a narrow area (e.g., answering a narrow research question).

- **Tool:** refers to any kind of tool support such as research prototypes or an implementation of one or more methods that support software engineering practices.

- **Metrics:** a type of contribution that provides guidelines or measures to assess different aspects of reliability.

- **Model:** a representation of information, a problem, or a topic instead of a way of solving a problem.

- **Evaluation:** has the same definition as mentioned in relation to research type (see Section 4.10.2). The contribution of evaluation research can be some sort of knowledge about evaluated techniques. If the main contribution of a study providing evaluation research is a tool, method or metric, such a study is classified under the specific category (as a tool, method or metric), but not as evaluation.

- **Review:** refers to the contribution type of secondary studies.

- **Benchmark:** refers to the studies that provide benchmarking datasets as their contribution.

- **Taxonomy:** provides classifications of gathered knowledge.

This category was added because some studies in our sample within the evaluation category also provided taxonomies as contributions.

### 4.10.4. Quality attributes

To address RQ2.4, we investigate how reliability was studied complying with the standard quality models in the context of mobile applications. For this purpose, we extracted (sub) quality attributes from each study and attempted to map the extracted QAs to QAs in the standard quality models ISO/IEC 25010:2011 and ISO/IEC 9126. Mapping to quality standards is important since standards play a key role in understanding whether proposed approaches are not only sound and efficient, but also likely to be economical and profitable for organizations. Demonstrations complying with standards show client organizations that their objectives and requirements are being achieved, with the potential for a better market position [39]. This way clients can assure that existing solutions can be applied effectively in daily industrial practice. We did not define the classification scheme prior for this RQ. Since the QAs are dependent on the context of the study, and terms used to refer to QAs are dependent on the author's intention, it is difficult to define a classification beforehand. Hence, we developed the classification during the data extraction process. The classification evolved and became stable with extracted data from the studies. The categorization of QAs is presented in Section 5.2.4.

### 4.10.5. Research method

In empirical software engineering research, there exists no consensus on how to classify research methods, which leads to various forms of classification (e.g., [62–66]). Furthermore, there are a range of ways to categorize the case study research method (e.g., [67–70]). As indicated by Journals and Wiley Online Library [51], it remains evident that standards of empirical research in other disciplines, such as social and medical sciences, are more mature compared to empirical research in software engineering, since such disciplines have a long history of empirical research [71]. Hence, we looked into a classification provided by Yin [72], who introduced five major research methods for social research: experiment, survey, archival analysis, history and case study. Yin further explained types of case studies based on the sources of evidence used to conduct research: documentation, archival records, interviews, direct observation, participant observation and physical artefacts.

In our review, not all selected studies state the research method that was used in their research. Hence, we defined the following categories to consistently classify empirical research in our study, which answers RQ3.1. First, we derived the main categories primarily based on the classification provided by Yin [72], and then we adapted definitions as follows to fit with relevant software engineering literature.

- **Experiment:** is a study in which an intervention is deliberately introduced to observe its effects [73]. These studies use an experimental method in a laboratory setting (i.e., not a natural setting, but manipulative setting) to empirically validate or evaluate their proposed solutions and approaches.

- **Survey:** is conducted to collect knowledge, attitudes, and ideas from a particular group of people using questionnaires or interviews.

- **Case study:** is conducted in a natural setting and is less controlled than an experimental setting. These studies can be a worked example or an investigation of real projects over a period to understand a certain phenomenon in-depth. These studies help to derive new hypotheses, theories, and test existing theories in different contexts. For example, case studies may involve tracking a particular attribute (e.g., number of bugs, amount of technical debt), establishing relationships between attributes, or building a model to predict bugs.

  – Archival research: We combined documentation and archival records from Yin [72] under this category, since the distinction of these two areas is not particularly relevant in software engineering research. Archival research is a type of case study that investigates historical data that are archived by companies which may require permission to access or are publicly available data such as from open-source repositories (e.g., GitHub, F-Droid for Android). Data can be collected from, e.g., meeting minutes, organizational documents such as policy documents, documents from different development phases such as design documents, requirement specifications, source code, test reports, organizational charts, financial records or previously recorded measurements such as effort and failure data [70,74].

  – Observational research: This indicates the same category, direct observation as in Yin [72]. A case study can be observational based on its data collection method. These types of studies typically use a qualitative data collection method where the researcher investigates (observes) software engineers while they perform their tasks in the work environment. With this method, non-verbal data such as participants' emotions, how they communicate with the team, how they complete tasks, what issues they deal with and other daily activities can be observed. Observation can be known or unknown to participants. Some strategies used in these types of research are monitoring software engineers with a video recorder or observing meetings where participants communicate, which may generate information.

  – Action research: This category is adapted from participant observation as in Yin [72]. Unlike observational research, in action research, researchers actively participate as a part of the organization and interact with participants at a higher level (i.e., the researcher is not an independent observer, but a participant) during the investigation.

- **Simulation research:** Simulation research helps with answering what-if questions (enabling observations by moving forward), while other methods permit answering the ''What happened, and how, and why?'' questions (i.e., looking backwards across history). Researchers use an executable model (i.e., transferred to executable source code from a mathematical/computational model) to represent a real-world entity. In software engineering, this entity may be a process, product, or person [62]. Researchers develop simulation models defining variables and their relationships and then run the model to analyse output variables and relationships among those variables for different input parameters. Simulations can avoid effort and set up costs for real-world experiments or may be convenient when resources are generally unavailable. Thus, they potentially help to reduce the problem of data unavailability in empirical research (e.g., when data is difficult or too costly to access), since simulation produces its own ''virtual'' data [74].

- **Review:** Secondary studies that analyse previous literature are classified as reviews.

### 4.10.6. Study settings

In an applied discipline such as software engineering, industrial involvement in research is extremely important. For many years, the need for collaboration between industry and academia in software engineering research has been acknowledged [75]. To answer RQ3.2, we investigate the study setting in which empirical studies are performed. Our classes are academic, industrial or both academic and industrial (mixed).

To classify studies based on setting, we examine to what extent industries have been involved in the selected studies. We found two types of study settings: studies from fully academic environments and studies that engage with industry, which, as noted above, we classify as a mixed

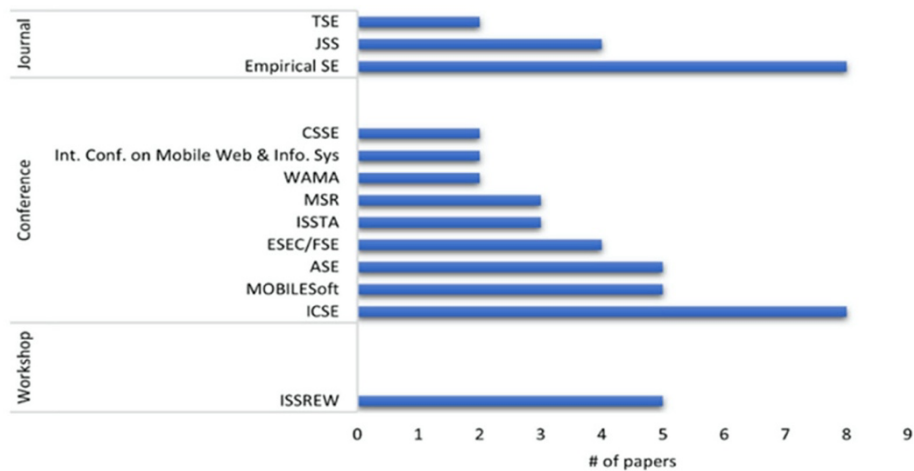
Figure 3. Venues that published more than one study on reliability of Android apps.

setting. To determine whether a study is industrial or mixed, we adapted two factors from a previous study [76]: involvement of practitioners, and use of datasets. Involvement of practitioners (who were actively employed in industry during the research) while performing the study can be through surveys or experiments. In addition, we also considered authors' affiliations. Author affiliations help to identify whether authors are from academia (employed with a university or research institute), industry (employed with a company) or from both (joint authorship). As explained by Garousi et al. [75], we believe that the involvement of practitioners in writing research papers enables both perspectives of academia and industry to be incorporated. Regarding datasets, some researchers use datasets from private companies, perhaps in addition to open-source datasets, which also indicates a level of collaboration with industry; thus these works are classified as mixed settings.

**4.10.7. Data**
To answer RQ3.3 we analysed details of the data used in empirical studies and classified studies based on three features of the data sources: (1) whether datasets are publicly available, (2) whether the studies analyse open-source systems or commercial systems, and (3) the type of artefacts used (e.g., bug reports, app source files) in the studies. Section 5.3.3 provides the classification of papers according to these categories.

## 5. RESULTS AND ANALYSIS

In this section, we present the results of the mapping study to answer our research questions. The Sections 5.1–5.3 are organized based on the research questions RQ1, RQ2 and RQ3 and their related sub-questions.

**5.1. RQ1: When and by whom has reliability of Android apps been studied?**
*5.1.1. Most active authors and venues (RQ1.1)*
Most authors have published three or fewer papers. Table 4 shows the most active authors in this field (i.e., 30 authors who published three or more papers) with their names and affiliations. The full list of papers including all the authors is provided in our online repository [43] (see Appendix B). Regarding publication venues, the selected 87 studies were published in 47 different venues (check Appendix C in Wimalasooriya et al. [43] for the full list of venues), including 57 conference papers, 24 journal articles and 6 workshop papers. The top eleven venues in which two or more studies were published are listed in. The total number of studies in this list is 53 which means around 34 other selected studies are not published in these venues. This result indicates that considering only the top venues would result in the omission of a number of related studies in systematic mapping studies. The results in Table 5 indicate that the journal of Empirical Software Engineering and the International Conference on Software Engineering (ICSE) published more than any other venues. We have plotted these results in Fig. 3.

*5.1.2. Research interest over time (RQ1.2)*
We analysed publication trends in terms of distribution of publications over the period covered in this mapping study, which is from 2008 to July 2021. As Fig. 4 shows, interest in this topic has increased in general. Even though the period of this mapping study is from the year 2008, research interest in operational reliability started around 2012. This may be due to various reasons such as (1) the first research focussed on the app store was published in 2010 [77] (which was two years after the first release of Android) and the fact that (2) mobile apps became relatively popular with significant increases in sales of smartphones in 2012 after the revolution of the touchscreen smartphone in 2010.

In Fig. 4, the first spike in research can be seen between 2013 and 2014. After 2014, at least 7 studies were published in each year. In 2018, the interest reached a peak, with 18 studies. This trend indicates that mobile app reliability is currently an active topic. It should be noted that our searches for this study were performed in July 2021, and thus, the figures for 2021 do not reflect studies published for the entire year.

**Summary outcomes for RQ1:** RQ1 explored the most active authors researching reliability of Android apps, venues in the area of Android app reliability and how research interest has changed over the time. Regarding authors, around 30 authors published three to four papers, while a large number of other authors published fewer papers. Considering the venues, the ICSE conference and the Empirical Software Engineering journal published the most in the area of reliability of Android apps research. However, the majority of studies are published in a number of other venues, albeit in fewer numbers. This means that considering only a list of the most active authors or venues is not sufficient to gain a broad understanding of the relevant research landscape. Furthermore, research interest on reliability of Android apps has increased in general over the years.

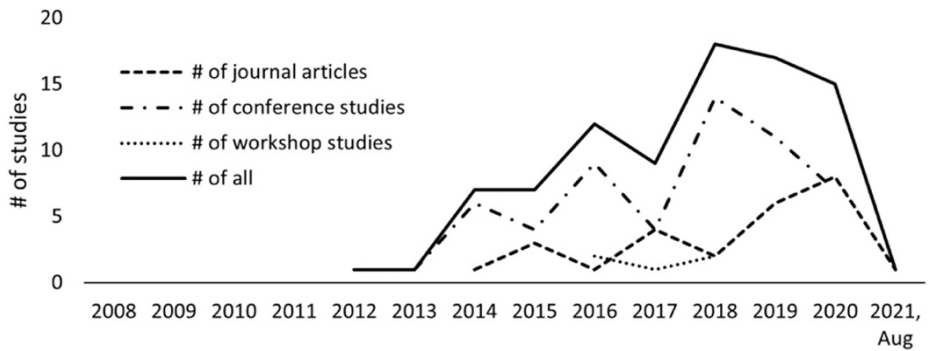

Figure 4. Publication trends by year for reliability of Android apps.

**5.2. RQ2: How has the reliability of Android apps been studied?**
*5.2.1. Research focus (RQ2.1)*
The selected studies were grouped into different categories according to the classification scheme described in Section 4.10 and Table 6 lists the relevant studies for each category of focus areas. The main topic categories used to structure the selected studies based on focus are: failure/crash analysis (35 studies), exception handling (21 studies), self-healing (8 studies), API-related issues (9 studies), ageing and rejuvenation (6 studies), context-awareness (4 studies), runtime change handling (4 studies) and other (6 studies). Below, we describe each category with the studies belonging to each group. If a study has more than one focus, the study is classified under multiple categories. For example, [s70] investigates runtime recovery via error handling and is grouped under both categories of runtime change- and exception-handling.

**Failure/crash analysis:** As Table 6 shows, it is noticeable that most of the studies published on the subject of Android app reliability were related to the category of failure/crash analysis (35 out of 87). The intent of these studies is to assist in the handling of operational failures, such as app crashes. These failures can be crashes which may cause an app to shut down, hang, restart, or other types of failure such as Application Not Responding (ANR) errors, phone freezes and phone self-reboots. Much research effort has also been committed to assist with testing and debugging approaches to detect crashes. Approaches in these works can be grouped into four types: (1) investigating the effectiveness of traditional testing (e.g., Monkey testing) in a mobile context to reveal app failures: [s1,s2,s14,s49]; (2) proposing new approaches to detect crashes: [s3], [s34,s41,s55,s60,s76]; (3) reproducing crashes to support the debugging process: [s22,s29, s30,s43,s51,s53,s54,s79]; and (4) analysing characteristics or root causes of crashes/failures: [s32,s65,s68]. Furthermore, two recent studies pointed out that Android Vitals provided in the Google console [s31] and program slicing for Android [s40] can facilitate current research on app testing. In addition to detecting crashes, fixing them was studied in [s47], [s52], and [s64]. The remaining studies focussed on predicting crashing releases [s77], failures caused by permission requests [s21], third party library updates [s35] and configuration errors due to developer mistakes [s37]. Studies [s15,s36] analyse the relationship between post-release faults (i.e., number of crashes) and app usage to derive a release quality measure (i.e., number of crashes per user), [s11] analyses modelling relationships between faults, failures and errors, [s5] surveys existing research on program analysis of Android apps, and [s83] investigates crash-inducing faults.

**Exception handling:** Most of the programming languages employed in mobile app development for Android apps use exception handling mechanisms to report and handle runtime failures (crashes). Properly using this mechanism assists developers to improve apps' ability to cope properly with runtime errors, thus, improving the fault-tolerance and robustness of apps. The second-most investigated area of the studies selected by this mapping study is exception handling (21 studies). Studies in this category investigated improving exception handling. For example, studies provided mechanisms to detect exception handling errors: focusing on null pointer exceptions [s48], Android system services [s71], or external resources [s69]. Another group of studies investigated relationships between uncaught exceptions and other factors such as: usage of Android abstractions [s38], evolution of exceptional and normal source code [s20], and extent of app usage (e.g., # of app users) [s15]. Studies Kechagia et al. [19],Kechagia and Spinellis [s59] examined how use of exceptions listed in Android API documentation can impact app quality. Furthermore, characterizing Android framework specific exceptions [s45]; exception fault localization [s45,s66]; proposing new mechanisms to handle exceptions [s39,s63]; and recommending correct exception handling code [s61] are among other investigations related to exception handling.

**Self-healing/recovery:** We grouped 8 studies into the category of self-healing abilities. Mobile apps are subject to high frequencies of user activities and configuration changes, such as switching between two apps, rotating the screen, resizing the screen, receiving phone calls that interrupts app execution, and so on. In such situations, Android apps must be able to handle many stops or restarts of apps without losing users' work or their interaction state. Hence, developers should implement the proper logic to save and restore the state of an app, as failing to do so may cause data loss, annoying usability issues, and even unexpected crashes.

Table 4. No. of papers published by the most active authors and their affiliations.

| Author | No. of papers | Institution | Country |
|---|---|---|---|
| Neamtiu, I. | 4 | New Jersey Institute of Technology, University of California | USA |
| Bissyandé, T. F. | 4 | University of Luxembourg | Luxembourg |
| Klein, J. | 4 | University of Luxembourg | Luxembourg |
| Li, L. | 4 | Monash University | Australia |
| Linares-Vasquez, M. | 4 | College of William & Mary, Universidad de los Andes | USA |
| Su, T. | 4 | Nanyang Technological University, East China Normal University, ETH Zurich | Singapore, China |
| Weng, C. | 4 | Wuhan University of Technology | China |
| Xiang, J. | 4 | Wuhan University of Technology | China |
| Zhao, D. | 4 | Wuhan University of Technology | China |
| Bernal-Cárdenas, C. | 3 | College of William & Mary | USA |
| Cacho, N. | 3 | Federal University of Rio Grande do Norte | Brazil |
| Sen Chen | 3 | East China Normal University, Nanyang Technological University, Tianjin University | China, Singapore |
| Gómez, M. | 3 | University of Lille | France |
| Jha, A. K. | 3 | Kyungpook National University | South Korea |
| Kechagia, M. | 3 | Athens University of Economics and Business, Delft University of Technology | Greece, Netherlands |
| Lee, W. J. | 3 | Kyungpook National University | South Korea |
| Lo, D. | 3 | Singapore Management University | Singapore |
| Mariani, L. | 3 | University of Milano — Bicocca | Italy |
| Micucci, D. | 3 | University of Milano — Bicocca | Italy |
| Oliveira, J. | 3 | Federal University of Rio Grande do Norte | Brazil |
| Poshyvanyk, D. | 3 | College of William & Mary | USA |
| Riganelli, O. | 3 | University of Milano — Bicocca | Italy |
| Rouvoy, R. | 3 | University of Lille | France |
| Spinellis, D | 3 | Athens University of Economics and Business | Greece |
| Fan, L. | 3 | East China Normal University, Nanyang Technological University, Nankai University | China, Singapore |
| Gao, J. | 3 | University of Luxembourg | Luxembourg |
| Paydar, S | 3 | Ferdowsi University of Mashhad | Iran |
| Pu, G. | 3 | East China Normal University | China |
| Su, Z. | 3 | University of California, ETH Zurich | USA |
| Xu, L. | 3 | New York University Shanghai, East China Normal University | China |

Table 5. Venues that published more than one study on the reliability of Android apps.

| Type | Venue | #studies |
|---|---|---|
| Conference (57 conference papers) | | |
| Conference | International Conference on Software Engineering (ICSE) | 8 |
| Conference | International Conference on Mobile Software Engineering and Systems (MOBILESoft) | 5 |
| Conference | International Conference on Automated Software Engineering (ASE) | 5 |
| Conference | Joint Meeting European Software Engineering Conference and Symposium on the Foundations of Software Engineering (ESEC/FSE) | 4 |
| Conference | International Symposium on Software Testing and Analysis (ISSTA) | 3 |
| Conference | International Working Conference on Mining Software Repositories (MSR) | 3 |
| Conference | International Workshop on App Market Analytics (WAMA) | 2 |
| Conference | International Conference on Mobile Web and Information Systems | 2 |
| Conference | International Conference on Computer Science and Software Engineering (CSSE) | 2 |
| Conference | *23 more conferences with 1 paper from each | |
| Journal (24 journal articles) | | |
| Journal | Empirical Software Engineering | 8 |
| Journal | Journal of Systems and Software | 4 |
| Journal | IEEE Transactions on Software Engineering | 2 |
| Journal | *9 more journals with one paper from each | |
| | Workshop (6 workshop papers) | |
| Workshop | International Symposium on Software Reliability Engineering Workshops, (ISSREW) | 5 |
| Workshop | *1 more workshop paper | |

Researchers proposed several solutions to help developers to facilitate apps' self-healing or recovery. Studies in this group focussed on detecting data loss errors, [s42,s62], providing benchmarks for data loss errors [s26], recovery mechanisms using different strategies, such as sealing off the crashing part of the app which allows the app to run with limited functionality rather than crashing in the wild [s58], or using exception handling mechanisms to recover from exception-related bugs [s70]. One study [s17] reveals that a common cause of runtime change issues is activity-restarting, and the authors proposed a restarting-free mechanism. Two studies proposed a mechanism to handle data persistence, [s44] proposed a technique to save/restore states' information which are stored in other variables beyond GUI elements (i.e., default implementation of apps events consider saving/restoring the states of GUI elements only), and [s75] aimed to assure constant access to remote data in case of network connectivity problems using a model-driven approach.

**API-related issues:** Researchers' next most addressed topic was API-related issues. These mainly address the deprecation and fault-proneness of APIs. Because of functionality evolution, as well as security and performance-related changes, APIs can become unnecessary and are no longer recommended for use (deprecated APIs). Thus, deprecated APIs need to be addressed, otherwise they may cause runtime crashes. We found 9 studies in which researchers were attempting to address these issues by characterizing deprecated Android APIs [s12], analysing causes for API failures [s33,s81], measuring the current practice of using the Android platform (taking inconsistencies between supported Android platform versions and API calls from a client app into account [s19,s72]), and also proposing approaches to automatically detect and correct API-related issues

Table 6. Research focus on reliability of Android apps and relevant studies.

| Research focus | Study ID | # of studies |
|---|---|---|
| Failure/Crash analysis | [s1,s2,s3,s5,s11,s14,s15,s21,s22,s29, s30,s31,s32,s34,s35,s37,s36,s40, s41,s43,s47,s49,s51,s52,s53,s54, s55,s60,s64,s65,s68,s76,s77,s79, s83] | 35 |
| Exception handling | [s13,s15,s18,s20,s38,s39,s45,s48, s50,s56,s59,s61,s63,s66,s69,s70, s71,s78,s84,s86,s87] | 21 |
| Self-healing/recovery | [s17,s26,s42,s44,s58,s62,s70,s75] | 8 |
| API-related issues | [s12,s16,s19,s33,s46,s57,s72,s81, s85] | 9 |
| Ageing and rejuvenation | [s4,s27,s28,s73,s74,s82] | 6 |
| Context-awareness | [s1,s6,s53,s54] | 4 |
| Runtime change handling | [s17,s24,s67,s80] | 4 |
| Other | [s7,s8,s9,s10,s23,s25] | 6 |

[s16,s85,s46]. Furthermore, [s57] investigated the impact of change- and fault-proneness of APIs.

**Ageing and rejuvenation:** Software rejuvenation (i.e., the manual or scheduled restart of an application) is a widely used approach to mitigate software ageing problems. Typical causes of software ageing are memory leaks, endless threads, storage fragmentation and unreleased locks [78]. Studies in this category describe how software ageing affects the reliability of Android apps and explore mechanisms to postpone software ageing-related failures using rejuvenation strategies. In [s74], the authors monitored available memory over time and found that ageing occurs in Android based on memory usage. The study also described that warm rejuvenation (application restart) has little effect on mitigating software ageing. Studies [s73,s27] introduced techniques to make optimal rejuvenation strategies: [s73] proposed starting rejuvenation when Android is in a critical state after estimating ageing point, and [s27] used modelling of both usage behaviour and the ageing process to study the impact of usage behaviour to rejuvenation. In [s28], predicting ageing indicators (system's free physical memory and application's heap memory) was investigated, and in [s4,s82], new rejuvenation strategies were proposed.

**Context-awareness:** Mobile apps are dependent on various external contextual factors such as network connectivity, hardware interruption, types of users, and so on. Due to frequently changing contextual factors, apps may not perform as expected in all conditions, causing various responsiveness and robustness issues. For example, apps running on a poor network connection may slow, show incorrect results or even crash. This category includes four studies that aim to improve apps' reliability by considering context-awareness [s1,s53,s54] by introducing techniques to reproduce context-sensitive testing, and [s6] discusses the state-of-the-art tools for automation testing of Android context-aware applications.

**Runtime change handling:** There are four studies that aim to handle runtime errors by focusing on reliable event handling of Android apps. These studies aimed to support the handling of runtime errors caused by asynchronous programming errors. Study [s24] explains how wide use of asynchrony can cause race errors that lead to use-after-free violations and proposed a tool to detect such errors. A use-after-free violation arises when a pointer is used when it is no longer pointed to any object (i.e., object is freed). When such errors are triggered, exceptions will occur. Improper handling of such exceptions may cause app crashes. Studies [s67,s80] also proposed new approaches to detect asynchronous errors, while [s17] proposed a new change-handling mechanism based on customized Android activity lifecycle states.

**Other:** As mentioned in the classification scheme, studies in this category do not have a specific focus on any of the more representative areas but are partially related to reliability. For example, [s7,s8] proposed quality models identifying the most important quality requirements for mobile apps. Study [s9] proposed a classification technique to extract non-functional requirements (including reliability-related issues) from user reviews, which gives insights about quality attributes of mobile apps from users' perspective.

*5.2.2. Research type (RQ2.2)*
Table 7 shows the studies that were classified according to the research types as defined in the classification scheme in Section 4.10.2. Of the total 87 studies, just over half (47 studies) were validation research, a further third (28 studies) were evaluation research and around one in ten (8 studies) were solution proposals. The experience report and opinion paper categories include one study for each type. Two secondary studies were classified as reviews, and we found no philosophical research. The vast majority (78 studies) of the selected studies are empirical and are highlighted in grey in Table 7. A detailed analysis of these empirical studies, including the empirical research methods and data used, is presented in Section 0. Solution proposal, philosophical research and opinion papers are the research types of non-empirical studies (see the classification in Section 4.10.2).

The bubble plot in Fig. 5 presents a visualization to provide further insights in answering this research question. The figure shows the distribution of the selected studies that were assessed in terms of their research focus, research type and contribution type. The size of the bubble indicates the relative number of studies belonging to that bubble. As seen in Fig. 5, handling crashes of mobile apps, which is the most focussed area, has been studied through different types of research, including evaluation (6 studies), validation (21), experience report (1), review study (1), and solution proposal (7). In the case of evaluation research, most of the evaluation was focused on improving reliability by handling crashes, exceptions and API-related issues. On the other hand, none of the approaches that concern context-awareness and runtime change (event handling) have been studied using evaluation research, and also only one evaluation research study for each of the self-healing and software rejuvenation topics is recorded. There are 4 studies that focus on context-awareness (see Table 6) with contributions (see the bubble plot in Fig. 5) of 3 tools comprising no evaluation research. This implies that there is a need for more evaluation research that could assess the effectiveness of the tools proposed by existing research.

Table 7. Type of research on reliability of Android apps.

| Research type | Study | # of studies | % of studies |
|---|---|---|---|
| Validation research | s4, s9, s15, s16, s17, s21, s22, s24, s27, s28, s30, s35, s37, s39, s40, s41, s42, s43, s46, s48, s51, s52, s53, s54, s55, s58, s61, s62, s63, s64, s66, s67, s68, s69, s70, s71, s73, s75, s76, s77, s79, s80, s82, s83, s84, s86, s87 | 47 | 54% |
| Evaluation research | s2, s7, s10, s12, s13, s14, s18, s19, s20, s23, s25, s26, s32, s33, s36, s38, s45, s47, s49, s50, s56, s57, s59, s72, s74, s78, s81, s85 | 28 | 32% |
| Experience report | s31 | 1 | 1% |
| Review | s5, s6 | 2 | 2% |
| Solution proposal | s1, s3, s11, s29, s34, s44, s60, s65 | 8 | 9% |
| Philosophical research | – | – | – |
| Opinion papers | s8 | 1 | 1% |

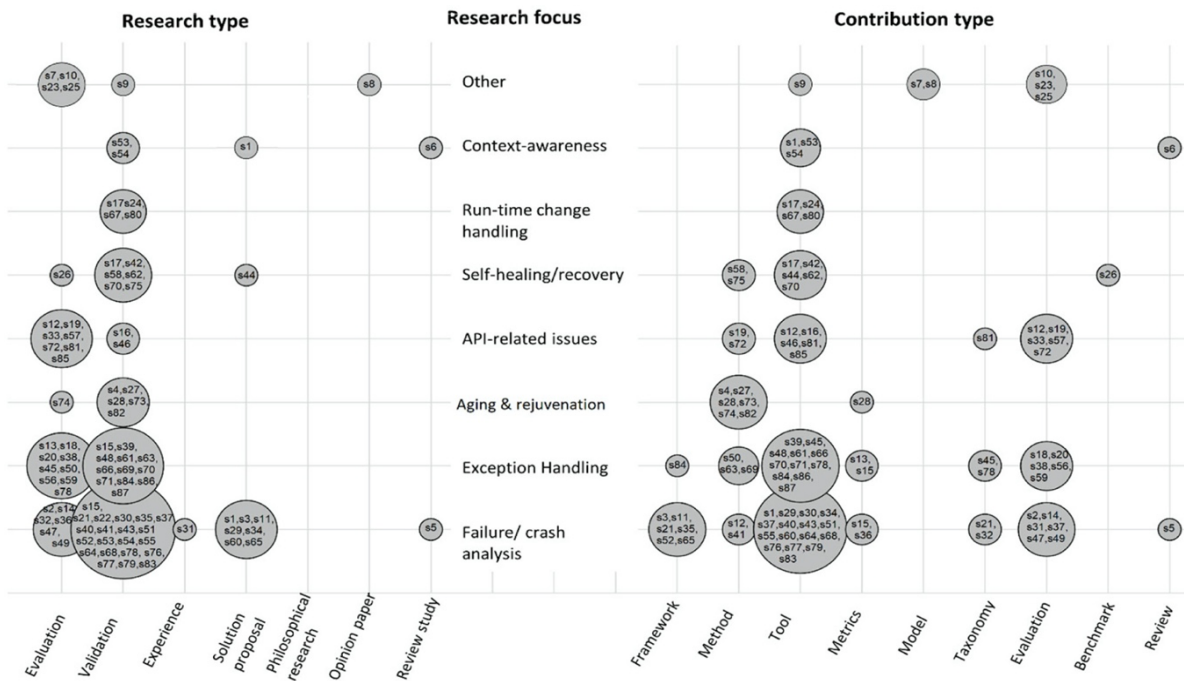

Figure 5. Research type, Research focus and Research contribution for reliability of Android apps.

*5.2.3. Research contribution (RQ2.3)*
Fig. 6 depicts the number of studies against each type of research contribution. The bubble plot in Fig. 5 also shows the contribution types of the studies against the research focus. As Fig. 6 shows, we found 42 studies that have contributed to tools by proposing new tools or contributing to an existing tool. Most of these tools are related to crash/failure analysis and exception handling. More details on these tools, such as the tool name and download links, can be found in our online repository [43] (see Appendix D).

The second-most common contribution type is evaluation, having the same definition as presented in Section 4.10.3. The reason why there is not the same number of evaluation studies in research type (see Table 7) and in contribution type (see Fig. 6) is that there are some studies where the major contribution is a tool, metric, method or taxonomy (although the studies include evaluation of an approach or technique) which were classified only under the respective category of its main contribution and not under the category of evaluation. If a study equally pays attention to two types of contribution, the IDs of such studies appear more than once in the bubble plot. There are 7 studies that provide frameworks, which are related to crash/failure analysis and exception handling. We observe that a relatively low number of studies offered metrics (4 studies). Taxonomy is usually a type of contribution resulting from evaluation research, so such studies also appear as evaluation type. Furthermore, if a study is concerned with representing information or providing an abstract classification, we classify them as models: [s7,s8] provide quality models for mobile apps. Review is the contribution type of the two secondary studies [s5,s6]. In the case of the provision of a benchmark, although there is other evaluation research that releases data to the public, we classify only [s26] as benchmark since the study's only contribution is providing a benchmark dataset for data loss bugs. The study presents their methodology and results for creating a public repository, without evidence of evaluation or any approaches or techniques.

*5.2.4. Quality attributes (RQ2.4)*
To answer this RQ, as we described in Section 4.10.4, we extracted QAs from the studies and mapped these to QAs in standard quality models. We observed that these QAs were not consistent with a uniform quality model. Some QAs are adopted from the standard models such as ISO/IEC 9126 and ISO/IEC 25010, or other standards (IEEE standard [79]). For example, robustness [79,80] is one such popular term used very often in the selected studies, even though it is not defined in the recent quality models.

Table 8. Quality attributes of reliability of Android apps.

| QA Sub-QA (as stated in the papers) | QA Sub-QA (as defined in the standard) | Studies | No. of studies |
|---|---|---|---|
| **(1) Reliability** | **Reliability** | [s6,s23,s25,s27,s28,s37,s39,s45,s46,s73,s74,s76,s78,s82, s84] | 15 |
| Availability | Availability | [s4,s27,s28,s37,s75,s82,s87] | 7 |
| Fault Tolerance/Robustness | Fault Tolerance | [s33,s39], [*Robustness*: [s14,s18,s20,s34,s38,s39,s45, s49,s53,s56,s57,s63,s69,s84]] | 15 |
| Recoverability (self-healing) | Recoverability | [s17,s26,s42,s44,s58,s62,s70] | 7 |
| | Maturity | None | – |
| **(2) Responsiveness** | **Performance efficiency** Time behaviour | [*Responsiveness*: [s2,s14,s17,s27,s28,s53,s68,s75,s80,s82]] | 10 |
| **(3) User experience/ Satisfaction** | **Quality in use** Satisfaction | [s3,s10], [*User experience*: [s3,s10,s12,s13,s16,s17,s21, s27,s33,s41,s61,s68,s73,s74,s75,s82]] | 16 |
| **(4) Other QA** | | | |
| Privacy/Security | Security | [*Privacy*: [s3]], [s12,s11,s18,s21,s23,s25,s37,s46,s71,s72] | 11 |
| Performance | Performance | [s1,s4,s12,s25,s75,s80] | 6 |
| Compatibility | Compatibility | [s12,s16,s18,s19,s25,s35,s46,s72,s81,s85] | 10 |
| Maintainability | Maintainability | [s23] | 1 |
| Usability | Usability | [s25] | 1 |
| **(5) QAs in general** | | [s7,s8,s9,s5,s78,s79,s86] | 7 |

Table 8 presents the extracted QAs from the studies, showing where and how often each QA was found. When different terminologies other than those defined in the ISO/IEC 25010 were used, we left relevant studies labelled with the terms as stated in the study. For example, studies that used ''Robustness'' as a sub-QA are presented with this label (see the ''Studies'' column). Some studies appear multiple times in one row because they use both QAs (as in standards) and other terminologies (labelled) interchangeably. For example, [s39] appear two times because the study used both Fault tolerance and Robustness.

As observed in Table 8, there are five types of studies in terms of QAs. First is the group of studies that target reliability and its sub-attributes. It is evident that papers explain their intention to improve reliability through specific sub-reliability attributes, so those papers appear against both reliability and its sub-attributes. Second and third, the attributes of responsiveness and user experience were very common terminologies used in the studies. These quality attributes can also be interrelated, though we distinguished the studies based on the authors' main concern. Nevertheless, some studies target multiple attributes simultaneously. For example, [s53] states that they aim to improve the robustness and responsiveness of apps against environmental interference.

The main concern of [s17] was to address poor responsiveness and recover user interaction state (i.e., recoverability) during runtime changes. Such studies appear against multiple QAs in Table 8. The fourth category refers to other QA. Papers in this category target reliability and discuss how reliability impacts other QAs or how other QAs cause reliability-related issues. For example, papers cover: how the evolution of APIs causes compatibility issues, which then result in the failed execution of apps [s12], or how security, backward compatibility and resource utilization are root causes of exceptions [s18]. Study [s23] showed that there are relationships between different types of bugs (reliability, security, maintainability) in mobile apps by analysing the correlations between them. The fifth category covers instances when researchers discuss all QAs in general, including reliability.

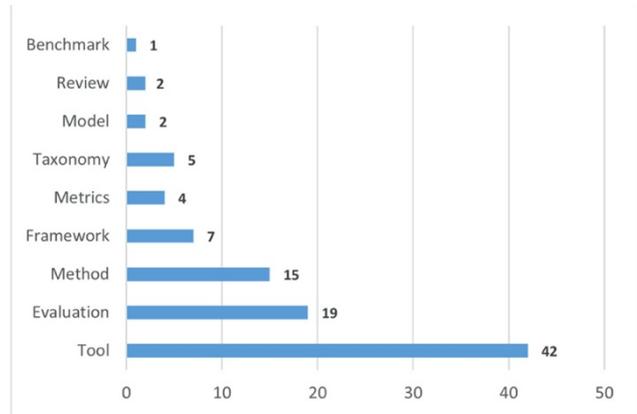

Figure 6. Research outcome on reliability of Android apps.

According to Table 8, we observe that the most frequently addressed reliability-related QAs are user experience (16 studies) and fault-tolerance/robustness (15 studies). Table 8 excludes 23 papers as they did not refer to any QAs. These papers addressed apps' runtime execution failures/crashes or exception handling without mentioning or linking these aspects to any particular quality attributes. Hence, we did not map such papers. On the other hand, some papers (e.g., [s6,s23,s25,s46]) only referred to the high-level attribute of reliability, ignoring detailed aspects and sub-attributes (e.g., availability, maturity).

### 5.2.5. Metrics/measures (RQ2.5)

Research addressing software metrics and measurements has been undertaken over several decades. As a result, several groups of metrics are available, across software processes, projects, and products [81]. However, previous

Table 9. Metrics and measures used to study reliability of Android apps.

| Metric/Measure type | Example for each type of Metrics and Measures from each study | No. of studies |
|---|---|---|
| Crashes/failures | $CrashRate_n$, $CrashDelay_n$, $FailureRate_n$, $FailureDelay_n$ [s14]; # *reliability bugs* [s23]; # *crashes* [s2,s39]; #crashes in the API [s57]; # ANR errors per app [s2,s14]; App response time [s4,s27]; # ANR, SNR events [s68], # bugs/crashes fixed in APIs [s57] | 9 |
| Exceptions | *Token Exceptions*: # potential bad token exception [s13]; *Exceptions*: # exceptions [s15,s36]; Uncaught Flow [s20,s38]; # uncaught exceptions [s39]; *Total # exceptions, # unique exceptions, # occurrences of the same exception, # occurrences of the same exception from the same class* [s50], # times that exception E occurs on code snippets that contains API method (m) [s61] | 8 |
| Responsiveness | App response time/App launch time [s4,s27,s28,s73], UI response time [s82] | 4 |
| User experience | *Complaint density* [s25]; % of issues reported in user reviews [s41] | 2 |

In addition to these metrics, researchers considered various types of other metrics which are indirectly related to reliability (environment- or resource utilization-related metrics such as metrics related to memory usage, CPU usage, app usage and API metrics) in addressing their research objectives.

studies have shown that traditional metrics and methods established in software engineering should be extended and/or re-assessed before applying them in a new domain, such as in the domain of mobile apps or web applications [82]. In order to deliver against different measurement needs and to accommodate the complexity and differences of mobile applications, a variety of metrics have been proposed, studied or validated in the context of mobile apps reliability, which we now discuss.

In Table 9, we present the metrics that refer to reliability attributes, the study/ies in which they were investigated, and the number of studies related to each metric. The metrics were used mainly to assess reliability-related attributes and to analyse relationships between reliability and other factors, such as software ageing, deprecated APIs, app usage, and so on. Based on the retrieved data, app failures/crashes (9 papers) or exception-related metrics (8 papers) are clearly the most studied metric categories.

Table 9 contains only the metrics that are directly related to reliability. The complete list, including the metrics related to other factors, is provided online (see Appendix E) [43]

**Summary outcomes for RQ2:** Most of the research on reliability of Android apps focused on handling crashes and exceptions. Other areas considered were API-related issues, self-healing, context-awareness, runtime change, ageing and rejuvenation. These research studies were mainly evaluation- or validation-type research, while only a few studies were in the form of solution proposals, experience reports, opinion papers and review studies. Contributions of these research works were mostly in the form of tools. In terms of quality attributes considered by researchers, reliability-related attributes such as availability, fault-tolerance, robustness and recoverability were considered. In addition to reliability attributes, there are other quality attributes such as responsiveness, user experience and compatibility that were also the focus of researchers to improve the reliability of Android apps. However, other studies (23 studies) did not refer to any types of quality attributes. Finally, on the metrics and measures used to study reliability of Android apps, the number of crashes and number of exceptions are the most common.

### 5.3. RQ3: Which studies have investigated the reliability of Android apps empirically?

As we mentioned in Section 5.2.2, we can divide the studies into either empirical or non-empirical research. When considering this split, 78 of the 87 studies represent empirical research, meaning just 9 are not empirical. We further analysed these empirical studies; thus, the rest of this section is focused on these 78 empirical studies. We investigated the research methods employed, the data used, and the environment (industrial/academic) in which these empirical investigations were performed. Thereafter, we explore the challenges and limitations of these studies.

*5.3.1. Research methods (RQ3.1)*
Table 10 reports the empirical studies from our sample of papers (78 studies) alongside the research methods used. These outcomes are also mapped to the research focus of the studies. Just 9 papers were not empirically based: [s1,s3,s8,s11,s29,s34, s44,s60,s65]; and these are excluded from the table.

As discussed in our classification in Section 4.10.5, we consider 5 main types of empirical methods: experiment, case study, simulation research, survey and review. Furthermore, due to there being numerous forms of case study, we looked into the incidence of three different types: archival, observational and action research. Note that some empirical studies use a combination of research methods, so such studies appear more than once in the table and are highlighted in grey. For example, [s18] conducted a mining study of stack traces from open-source Android projects (case study) and an exploratory survey with app developers to obtain a thorough understanding of common exception handling bugs that they face. Again, studies with more than one research focus appear under multiple research focus areas in Table 10.

'Totals' are the number of unique paper Ids for each empirical method type and each focus area. The majority of the empirical studies (54 of 78) used experiments, with the next most frequently used empirical method being archival based case studies (in 24 studies). Only one of the selected studies used an observational approach, with no study in our cohort of papers using action research. Six (6) studies were recorded for simulation research and two/three for each of the other methods: surveys and reviews. An important observation here is the relative lack of surveys, action or observational research, even though such research methods allow the community to investigate reliability in real-world contexts rather than in experimental settings.

In addition, considering the research areas in focus in empirical studies, all such areas were studied through at least one empirical study. As Fig. 7 shows, reliability issues related to context awareness issues (3 studies), self-healing ability (4 studies), software ageing and rejuvenation (6 studies) and runtime-change handling (7 studies) are the least empirically investigated areas among the selected studies.

Table 10. Empirical research methods used for research focussed on reliability of Android apps.*(Continued on next row)*

| Research method | Failure/crash analysis | Exception handling | Aging & rejuvenation | API-related issues | Self-healing/recovery | Run-time change handling | Context-awareness | Other | Total |
|---|---|---|---|---|---|---|---|---|---|
| Experiment | s2, s14, s15, s22, s30, s35, s40, s41, s43, s47, s49, s51, s52, s54, s55, s64, s68, s76, s77, s79, s83 | s15, s20, s38, s39, s48, s56, s61, s63, s66, s69, s70, s71, s84, s86, s87 | s4, s27, s28, s73, s74, s82 | s16, s81, s85 | s17, s58, s70, s75 | s17, s26, s24, s42, s62, s67, s80 | s54 | s23 | 54 |
| Case study |  |  |  |  |  |  |  |  |  |
| – Archival Research | s13, s21, s32, s37 | s18, s45, s50, s56, s59, s69, s78, s87 |  | s12, s19, s33, s46, s57, s72, s81 |  | s26 |  | s7, s9, s10, s25 | 24 |
| – Observational | s31 |  |  |  |  |  |  |  | 1 |

Table 10. *(Continued).*

| Research method | Failure/crash analysis | Exception handling | Aging & rejuvenation | API-related issues | Self-healing/recovery | Run-time change handling | Context-awareness | Other | Total |
|---|---|---|---|---|---|---|---|---|---|
| –Action research |  |  |  |  |  |  |  |  |  |
| Simulation research | s36, s53 | s15 | s73, s74, s82 |  |  |  | s53 |  | 6 |
| Survey |  | s18, s78 |  | s57 |  |  |  |  | 3 |
| Review | s5 |  |  |  |  |  | s6 |  | 2 |
| Total | 29 | 20 | 6 | 9 | 4 | 7 | 3 | 5 |  |

*5.3.2. Study settings (RQ3.2)*
Table 11 shows the results of the distribution of study settings. According to the classification scheme described in Section 4.10.6, we identified two types of studies among our selected studies: academic and mixed (both academic and industry).

Based on our observation, most of the empirical studies (64 out of 78) are from academic settings. Studies that were performed in an academic setting (refer to Section 4.10.6) without any industrial collaboration, or where the research environment was not described sufficiently (e.g., when studies do not mention where the data come from), were classified as academic. A relatively small proportion of studies, i.e., 14 studies, were from a mixed setting. These 14 studies collaborated with industry using one or more of the techniques described in our classification scheme (Section 4.10.6), including joint authorship [s11,s24,s48, s50,s68], experiment/surveys with practitioners [s18,s31,s56,s57, s75,s78,s87], or the use of industrial data/databases [s31,s33,s48, s50].

*5.3.3. Study information/data (RQ3.3)*
In order to answer this RQ, we gathered information about the data that was used in the empirical studies. Some studies collect their own data from online repositories such as F-Droid, GitHub and Google Play, while others use datasets from previously published literature. Table 12 reports the types of apps or projects (i.e., OSS, commercial) and the specific artefacts that were studied (e.g., source files, commits) and the availability of datasets.

Not all studies release their data to the public as Table 12 shows, which restricts other researchers from performing replication research. Most studies analysed OSS (only 28% of the studies analysed commercial software.) data across a range of artefact types. As shown in Table 12, these artefacts are crash and bug reports (e.g., bug reports automatically generated by testing tools, stack traces), apps source files (e.g., apps' source code, Android-Manifest.xml files and comments), apps' binaries (e.g., apk files), the Android framework/API, the Android OS, commits by developers, Android/API documentation, app reviews and developer forums or blogs (e.g., Stack Overflow, developer discussions in issue trackers). A few studies (15 studies in total) used other types of artefacts (see the last column of artefacts in Table 12) such as release notes [s10,s70], app usage data from the Google Analytics platform [s15,s36], memory usage data [s28,s74], and app store publishing policies [s7].

*5.3.4. Research gaps and limitations (RQ3.4)*
To answer RQ3.4, we extracted research gaps and limitations in two ways: (1) by extracting these from selected studies as noted by the respective authors and, (2) based on the frequencies returned from the classifications.

*Tool support:* The lack of tool support for predicting app crashes or failures is highlighted in three studies [s15], [s61], [s77]. Study [s15] shows that app usage is highly important in terms of accurately predicting exceptions and the authors advised the community to incorporate usage data in prediction models. Furthermore, the authors in [s15] noted the necessity of tool support to detect API exceptions (e.g., documented unchecked exceptions). Five other studies also mentioned the need for tools to handle API-related issues. As argued in [s12], tools are required to fix the use of deprecated APIs in the wild, and the study provided a large dataset of apps that could help the community to systematically learn patterns for fixing deprecated APIs. Study [s16] attempted to automatically detect API-related compatibility issues, although they acknowledged the need for enhancements to handle forward compatibility issues. Moreover, tools are required to detect crashes caused by incorrect resource handling and API permission handling [s32], to detect asynchronous program errors with complicated exceptions such as 'file not found' and database corruption [s67], and for localizing framework-specific exceptions, since existing tools support only limited types of exceptions [s45].

In addition, the first program slicing approach for Android was introduced in [s40]. The study proposed a tool for slicing Android apps and shows how slicing can facilitate

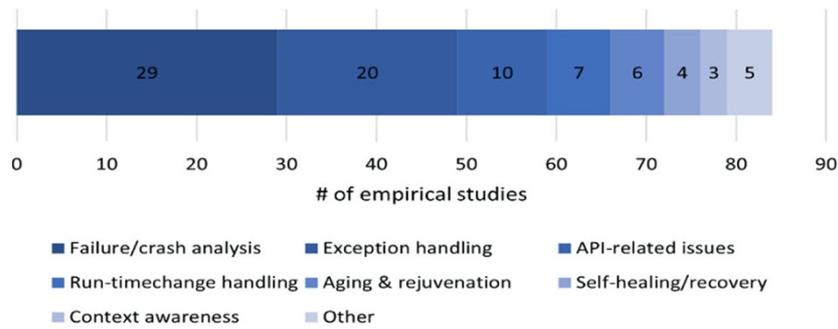

Figure 7. Empirical studies for research focussed on reliability of Android apps.

Table 11. Study settings for research focussed on reliability of Android apps.

| Study setting | Frequency | Percentage |
|---|---|---|
| Academic | 64 studies | 82% |
| Academic + Industry (Mixed) | 14 studies [s9,s18,s24,s31,s33,s48, s50,s56,s57,s68,s69,s75,s78,s87] | 18% |

fault localization, testing and debugging in the mobile domain. Since this is the only study that explores program slicing in a mobile context, more studies are required to validate the effectiveness of slicing

*Integrate user feedback into development:* Recent studies have suggested that mining user reviews can help with improving approaches to user acceptance testing since reviews contains information related to problems faced by users, and research in this area is beginning to grow [83]. In [s41], the authors proposed an automated approach to link user reviews to stack traces. They also suggested future research directions towards summarizing reviews linked with stack traces, prioritizing bugs by taking user reviews into account, and generating test cases directly from user reviews. These could help developers when fixing bugs and could complement current testing approaches. These research directions have also been mentioned in other studies (e.g., [83,84]).

*Relationships between quality attributes*: Another important challenge in maintaining the quality of software more generally is, that quality attributes can be interrelated, potentially impeding the clear identification of issues. Questions may arise about what quality issues are to be fixed first, if fixing one affects others. Hence, there is no single aspect of quality that is more important than others when software is complex, distributed and integrated. In accordance with the previously discussed findings in Section 5.2.4, we found just one study that investigated the relationships or trade-offs between reliability and other QAs (i.e., [s23]), where correlations between security, maintainability and reliability bugs were analysed. Future work aimed at providing tools support to analyse the relationships among multiple quality attributes while performing maintenance activities, and research addressing quality trade-offs, will be important to achieving the optimal balance between multiple quality attributes.

*Following standard quality models:* As we described in Section 5.2.4, not all researchers refer to international quality standards in their work. Since reliability is defined as a quality attribute in standard quality models (e.g., ISO/IEC 9126, ISO/IEC 25010), following standards or taking details of sub-attributes (e.g., availability, recoverability) into account would have been more useful since standards provide a level of assurance of the effectiveness of any proposed approaches in the real world (refer to Section 4.10.4).

*App ecosystem-related issues:* The continuous evolution of the app ecosystem is a major challenge to running apps reliably in the wild. Even though apps may well be released with good quality after proper testing, they may not function properly in different environments (different versions of OSs, different users). Many crashes and other runtime issues are related to the ecosystem itself. Therefore, researchers need to pay more attention to ecosystem issues. Previous studies have proposed solutions to dealing with new versions of the app store [s3], also noting the need for enhancements to handle practical limitations. Another recent study [s68] explained that more ecosystem research is needed to focus on issues such as inter-app and app-OS interactions, rather than testing that focuses on a single application at a time.

*Misunderstanding/limitations of Android documentation:* There are a number of mechanisms related to API exception handling that are not understood properly by developers [1,18]. For example, Android documentation provides conditional checks, such as checks on API parameters, activity states (e.g., IllegalState) and SDK versions. Adding appropriate conditional checks can avoid exceptions from Parameter Error, Lifecycle Error, Resource Error and the like. In addition, APIs may evolve fast and developers can become confused in such situations, which may lead to the misuse of APIs [18]. To limit Android-specific exceptions, developers need to gain and sustain a full understanding of the Android system. Future research on a range of supporting tools that provide automatic rule violations and patch suggestions based on Android API documentation will be beneficial. Such tools could be used as plugins to assist developers during development.

In addition, Android documentation has its own limitations, such as not clearly illustrating some error-handling strategies. For example, Android documentation does not provide a full list of system permissions [85]. Developers are more likely to avoid mistakes if there is proper documentation for Android [s37]. As argued in a recent review [s5], more research is required that systematically collects and characterizes such error-handling mechanisms, and they go on to provide recommendations to formalize Android documentation.

*Lack of evaluation research:* Based on our analysis, the majority of mobile app reliability studies are validation type research. In contrast, there were few or no evaluation studies found among those that proposed approaches to improve reliability in terms of context-awareness, runtime change (event) handling, self-healing and software rejuvenation (see Section 5.2.2). These areas are the least empirically investigated areas among the selected studies. Hence, future studies need to systematically evaluate these approaches in more complex and real-world projects to identify real challenges, advances and opportunities.

Table 12. Data used in empirical investigation for research focussed on reliability of Android apps.

| Paper Id | Dataset publicly available? | OSS | Commercial | Issue/crash reports | Apps (source files) | Apps (in binary level) | Android framework | Android OS | Commits | Android documentation | App reviews | Forums/Developer discussions | Other |
|---|---|---|---|---|---|---|---|---|---|---|---|---|---|
| [s2] | | ✓ | | | | ✓ | | | | | | | |
| [s4] | | ✓ | | | | ✓ | | | | | | | |
| [s5] | | | | | | | | | | | | | |
| [s6] | | | | | | | | | | | | | |
| [s7] | | | | | | | | | | | | | |
| [s9] | ✓ | ✓ | | | | | | | | | ✓ | | |
| [s10] | | ✓ | | | | ✓ | | | ✓ | | ✓ | | |
| [s12] | ✓ | ✓ | | | | ✓ | ✓ | | | | | | |
| [s13] | | ✓ | | | | ✓ | | | | | | | |
| [s14] | | ✓ | | | | ✓ | | | | | | | |
| [s15] | ✓ | ✓ | ✓ | | | | | | | | | | |
| [s16] | ✓ | ✓ | ✓ | | | | ✓ | | | ✓ | | | |
| [s17] | | ✓ | | ✓ | ✓ | | ✓ | | | | | | |
| [s18] | ✓ | ✓ | | ✓ | ✓ | | | | | | | | |
| [s19] | | ✓ | | | | ✓ | | | ✓ | | | | |
| [s20] | | ✓ | | | ✓ | ✓ | | | | | | | |
| [s21] | ✓ | ✓ | ✓ | | | | | | ✓ | ✓ | ✓ | | |
| [s22] | | | ✓ | | | ✓ | | | | | | | |
| [s23] | | ✓ | | | ✓ | | | | | | | | |
| [s24] | | ✓ | | | | ✓ | | ✓ | | | | | |
| [s25] | | | | | | | | | | | ✓ | | |
| [s26] | | ✓ | | ✓ | | | | ✓ | | | | | |
| [s27] | | | | | | | ✓ | ✓ | | | | | |
| [s28] | | ✓ | | | | ✓ | | ✓ | | | | | |
| [s30] | ✓ | ✓ | | ✓ | | ✓ | | | | | | | |
| [s31] | | ✓ | ✓ | ✓ | | | | | | | | | |
| [s32] | ✓ | ✓ | | ✓ | ✓ | | | | | | | ✓ | |
| [s33] | | ✓ | ✓ | ✓ | | | ✓ | | | | | | |
| [s35] | ✓ | ✓ | | | | ✓ | | | | | | | |
| [s36] | ✓ | ✓ | ✓ | | | | | | | | | | |
| [s37] | | ✓ | | | ✓ | | | | | | | | |
| [s38] | | ✓ | | | ✓ | | | | | | | | |
| [s39] | | ✓ | ✓ | | ✓ | | | | | | | | |
| [s40] | | ✓ | ✓ | | | | | | | | | | |
| [s41] | ✓ | ✓ | | ✓ | | ✓ | | | | | ✓ | | |
| [s42] | ✓ | ✓ | | | ✓ | ✓ | | | | | | | |
| [s43] | ✓ | ✓ | | | | ✓ | | | | | | | |
| [s45] | ✓ | ✓ | | ✓ | ✓ | | | | | ✓ | | ✓ | |
| [s46] | | ✓ | | | ✓ | | | | | ✓ | | | |

| Paper Id | Dataset publicly available? | OSS | Commercial | Issue/crash reports | Apps (source files) | Apps (in binary level) | Android framework | Android OS | Commits | Android documentation | App reviews | Forums/Developer discussions | Other |
|---|---|---|---|---|---|---|---|---|---|---|---|---|---|
| [s47] | ✓ | ✓ | | | | ✓ | | | | | | | |
| [s48] | ✓ | ✓ | ✓ | | | | | | | | | | |
| [s49] | | ✓ | ✓ | | | ✓ | | | | | | | |
| [s50] | | ✓ | | ✓ | | | | | | | | | |
| [s51] | ✓ | ✓ | | | | ✓ | | | | | | | |
| [s52] | | ✓ | | ✓ | | | | | | | ✓ | | |
| [s53] | | ✓ | ✓ | | | ✓ | | | | | | | |
| [s54] | | ✓ | ✓ | ✓ | | | | | | | | | |
| [s55] | | ✓ | | | | ✓ | | | | | | | |
| [s56] | | ✓ | ✓ | ✓ | | ✓ | ✓ | | | | | | |
| [s57] | ✓ | ✓ | | | | ✓ | | | | | ✓ | | |
| [s58] | | ✓ | ✓ | | | ✓ | | | | | | | |
| [s59] | | ✓ | ✓ | ✓ | | | ✓ | | | ✓ | | | |
| [s61] | ✓ | ✓ | | | ✓ | ✓ | | | | | | | |
| [s62] | ✓ | ✓ | | | | | | | | | | | |
| [s63] | | ✓ | ✓ | | | ✓ | | | | | | | |
| [s64] | | ✓ | | ✓ | | | | | | | | | |
| [s66] | | ✓ | | | | ✓ | | | | | | | |
| [s67] | | ✓ | | | | ✓ | ✓ | | | ✓ | | ✓ | |
| [s68] | ✓ | | | | | | | ✓ | | | | | |
| [s69] | | ✓ | | ✓ | | | | | | | | | |
| [s70] | | ✓ | | ✓ | | | | | | | | | ✓ |
| [s71] | | ✓ | | | | | ✓ | | | | | | |
| [s72] | ✓ | ✓ | ✓ | | | ✓ | | | | | ✓ | | |
| [s73] | | | ✓ | | | ✓ | | | | | | | |
| [s74] | | | ✓ | | | ✓ | | | | | | | |
| [s75] | | | | | | | ✓ | | | | | | |
| [s76] | | | | | | | ✓ | | | | | | |
| [s77] | | ✓ | | ✓ | ✓ | | | ✓ | | | ✓ | | |
| [s78] | ✓ | ✓ | ✓ | ✓ | ✓ | ✓ | | | | | | | |
| [s79] | ✓ | ✓ | | | | | | | | | ✓ | | |
| [s80] | ✓ | ✓ | ✓ | ✓ | ✓ | | | | | ✓ | | ✓ | |
| [s81] | ✓ | ✓ | | | | | | | | | | | ✓ |
| [s82] | | ✓ | | | | ✓ | | | | | | | |
| [s83] | ✓ | ✓ | | ✓ | | ✓ | | | | | ✓ | ✓ | |
| [s84] | | ✓ | | | | | | | | | | | |
| [s85] | | ✓ | | | ✓ | ✓ | | ✓ | | | | | |
| [s86] | | ✓ | | | | ✓ | ✓ | | | | | | |
| [s87] | ✓ | | | ✓ | | | | | ✓ | | | ✓ | |

*Generalizability and industry adoption:* We observed (see Section 5.3.1) that two-thirds of the empirical studies used experimental methods to conduct their research. Also, the majority of the proposed solutions were verified in an academic setting (see Section 5.3.2) using OSS projects, rather than in an industrial setting, which indicates that current approaches and empirical observations might not be generalizable to commercial software systems. As we also noted previously, research that uses empirical methods such as action research, observational research or surveys with practitioners allows the academic community to work closely with industry and provides opportunities to access company data repositories. More research of this nature would help with industrial-level validation and would inform our understanding of how to support methods' adoption, in transferring proposed approaches developed in the laboratory to real life practice, and vice versa.

**Summary outcomes for RQ3:** RQ3 further investigated a subset of the studies (78 out of 87) from RQ2, i.e., the empirical studies. The majority of these studies conducted experiments, while the next most used method is the case study. Almost all case studies are archival research. Only a few papers used other research methods, such as simulation, surveys, and reviews. In terms of study settings, 82% of the studies were conducted in an academic environment without any form of collaboration with industry. This RQ also provides information about the data sources used in the studies. The majority of the studies used OSS, while only 28% of the studies analysed commercial software. The most commonly used artefacts are issue reports and apps (in source-code and binary forms). The major limitations of existing research are neglecting quality standards, lack of evaluation research with industry and lack of tool support to detect and handle field failures early.

# 6. SUMMARY DISCUSSION AND IMPLICATIONS

The intent of this systematic mapping study was to collect, analyse and interpret all existing evidence related to the operational reliability of Android mobile apps. To the best of our knowledge there is no previous systematic review or mapping study performed in this area. We built a set of classification schemes to classify existing literature, in turn enabling us to identify research gaps to inform future research directions. We organized this mapping study based on three main RQs.

*RQ1. When and by whom has reliability of Android apps been studied?* In addressing RQ1, research trends in terms of time, venue, publication type and authors were analysed. Interest in this topic has increased in general since 2008 and the evidence suggests that the reliability of mobile apps is an active topic currently. Identification of the more frequently published authors and venues regarding this topic provides an indication to researchers as to where to look for similar research. However, the selected publications were drawn from a broad range of venues (47 venues), and nearly half of the selected studies were not published in the top venues list such as the International Conference on Software Engineering (ICSE) and Empirical Software Engineering (see Table 5). Therefore, we would not recommend limiting reliability-related literature reviews to consider only the top venues since they could miss a notable number of related studies.

*RQ2. How has the reliability of Android apps been studied?* Regarding RQ2, we analysed how reliability has been studied. We found seven different focus areas that researchers have tended to work on when seeking to improve reliability: failures/crashes, exception handling, self-healing ability/recoverability, API-related issues, ageing and rejuvenation, context awareness, runtime change/ event handling and others. Failures and exception handling are the most studied areas when considering the reliability of Android apps. Very little attention has been paid to reliability with respect to handling runtime changes (e.g., rotate screen, switch apps), context-awareness issues, and software rejuvenation. In terms of contribution types, most of the research reported to date has contributed to tool support. Most of these tools are developed to handle crashes/failures, exceptions or API-related issues (refer to Section 5.2.3). However, these tools still have several limitations as was explained in the previous section (Section 5.3.4). We also looked into the reliability of Android apps based on quality standards. It is important that researchers consider international standards since standards play a key role in achieving real-world goals [39]. We noticed that not all papers follow quality standards or pay attention to QAs as defined in standards such as ISO/IEC 9126 and ISO/IEC 25010 (refer to Section 5.2.4). In most cases, studies do not give explicit attention to any QAs in their research or they may refer to only the high-level attribute, which is 'reliability' in our case. Little evidence exists for analysis that goes beyond the high-level or that considers the sub-attributes (i.e., availability, maturity, recoverability, and fault tolerance) of reliability. Following the standards could have enabled the studies to assess reliability without missing its more detailed aspects.

In answering RQ2, we also analysed what metrics and measures have been used to investigate reliability. The most used metrics are based on failure data (e.g., # crashes per app, # uncaught exceptions per app, refer to Section 5.2.5). Furthermore, we found that more than half of the selected studies are validation research. More evaluation research is required, especially in relation to context-awareness, runtime change/event handling, software rejuvenation and self-healing, in order to validate existing tools, metrics and other approaches in real-world contexts to identify practical challenges and the effectiveness of these solutions.

*RQ3. Which studies have investigated the reliability of Android apps empirically?* Regarding RQ3, our objective was to further analyse the empirical studies to understand the methods, study settings, information/data sources and limitations and gaps of empirical research that had focussed on the reliability of Android apps. Of the 87 studies selected, 78 were empirical research. Among these studies, most employed experimental methods, followed by the archival-based case study method (used by close to one in three studies). Even though the case study approach enables researchers to conduct research in natural settings when compared to experimental methods, almost all of the case studies encountered in our sample focussed on archival-based research. Thus, the findings of these research studies are based on historical data. Historical data on its own may not be sufficient to enable strong conclusions to be drawn [86]; instead, such data might uncover useful but potentially dated information and patterns such as correlations [86]. Therefore, such findings must be tested more fully in current real world contexts [86], in order to reveal contemporary industrial challenges, since realism often add extra concerns [70].

Our results also showed that fewer than one in five empirical studies involved collaboration with industry in the conduct of the research. Thus, based on our results, we contend that there is a shortage of empirical research examining the actual use of reliability-related approaches – tools, methods – in relation to contemporary real-life problem scenarios and investigations. This could be remedied by researchers conducting more surveys (i.e., developers can report on the current situation and challenges when using proposed approaches), and also studies using action or observational research (i.e., researchers can report on what practitioners actually do as against what they say they do). In fact, the most realistic research setting is facilitated by action research [87]. Our results also provided information about the kinds of data and artefacts (e.g., source files, stack traces, APIs) used by previous researchers and if that data is made publicly available. This should be useful for researchers who are interested in continuing research in the area, and especially those performing replication studies. Practitioners may also gain knowledge about what approaches work well in particular work contexts.

Furthermore, beyond the above research gaps, we found several other research gaps and limitations based on the findings of our selected studies, which we described in the previous section (Section 5.3.4). These are the limitations regarding tool support, integration of user feedback into the app development process, consideration of trade-offs/relationships between reliability and other QAs, and addressing app ecosystem issues and limitations of Android documentation.

## 7. THREATS TO VALIDITY

According to Petersen et al. [42], descriptive validity, theoretical validity, interpretive validity, repeatability and generalizability should be addressed when evaluating mapping studies. We now consider the threats to validity that are relevant to this particular mapping study.

*Descriptive validity* is concerned with researchers' bias in the study selection and data extraction process [42]. To mitigate this risk, we developed a protocol which was reviewed by each author of this study, and we piloted the protocol against a test set. The test set included studies that we found by manually searching a carefully selected set of known relevant venues. The manual search was performed independently by two researchers and results were compared using inter-rater agreement, recording good agreement between the researchers. We resolved all disagreements through extensive discussions, while fine-tuning the protocol, including the study selection criteria. Regarding data extraction, there is a possibility that results can be biased since data extraction was performed by a single researcher with reliability checks performed by a second author. To verify data extraction, another author randomly selected 15% of the selected studies and completed data extraction. We held discussions in meetings until consensus was reached among the authors when there were disagreements, and this informed or resulted in revised extraction processes. Thus, we are confident that we have limited the effect of this threat.

*Theoretical validity* is related to the completeness of the set of selected studies [42], since there is a chance that we could have missed relevant studies. To minimize this threat, we tested several search strings against the test set and chose the one that returned the maximum number of studies in the test set. We also searched in a wide range of major software engineering databases (5 databases) to collect all relevant studies. Additionally, we collected more studies through backward and forward snowballing processes. Therefore, we were unlikely to have missed relevant studies, and so this is unlikely to affect our study findings and conclusion.

*Interpretive validity* reflects conclusion validity [42]. Interpreting data and drawing conclusions from the results may have been affected by the structure of the classification schemes [20]. To mitigate this threat, we followed existing guidelines and other systematic secondary studies to identify the categories in our classification schemes (e.g., [37,60,61,72]). Second, each author independently reviewed the classification schemes before applying them. Furthermore, we believe that slight misclassification would not change the main conclusions drawn in our study. We should also note that our analyses used frequencies of studies, following the convention in mapping studies. Just looking at frequencies of studies may not provide a complete understanding of a topic or domain, since research that occurs regularly may not necessarily be more important or more mature than infrequent or emergent research.

*Repeatability* of this study is achieved by the provision of a protocol which is extensively described in a way that can be easily replicated by other researchers. There is a possibility that another researcher may include a study that we excluded if it was published outside our search period. However, while this may change the actual numbers of publications, it is unlikely to change the overall results.

*Generalizability* deals with external validity [41,42], which reflects the extent to which the conclusions are applicable outside this particular study. Since this study considered only Android apps' reliability, our conclusions are only concerned with this specific context. Therefore, external validity threats may exist if the findings of the work are assessed against the wider body of evidence on reliability or if the conclusions drawn are extrapolated beyond the Android app context.

## 8. CONCLUSION

Our study analyses a total of 87 studies published between 2008 and 2021 in the area of Android app reliability. We provide classification schemes to structure the selected studies according to research focus, research type, research method, contributions, study settings and quality attributes, as considered in the studies. The metrics, datasets and artefacts used in the studies are also captured and summarized. We now present the conclusions of our study.

The selected studies have been published across 47 different venues. Therefore, looking only into the dominant venues or at the writings of most popular authors would not be sufficient to gain a full understanding of this topic. Most of the research (82%) has been conducted in academic environments and so there is a gap between the researched approaches and potential adaptation, in practice. Further research is required in which researchers should collaboratively work with practitioners. In doing so the researchers would do well to follow international standards (e.g., ISO/IEC 25010) to lend assurance of the effectiveness of the approaches being assessed.

Our classification of the work to date showed that researchers are more interested in dealing with crashes and exceptions to improve reliability, while there has been less of a research focus on enhancing reliability via self-healing abilities, addressing of API-related issues, adequate context awareness techniques, handling asynchronous programming errors and software rejuvenation. Our study summarized several research gaps: lack of tool support (for detecting and avoiding specific types of crashes and framework-specific exceptions, and for handling deprecated APIs), lack of evaluation research, the need to better integrate user feedback into the development process, broader app ecosystem-related issues and challenges, and to improve Android documentation.

**Future research directions:** There is room for future research to improve existing approaches. In terms of context-aware approaches, current techniques support only the apps that use low-level contexts (context data that are directly collected from sensors such as time, GPS, noise). Future research may work towards supporting high-level context data (e.g., apps that track movements deal with high-level contexts such as ''moving'' and ''resting'' which are combinations of lower-level contexts). With regard to software rejuvenation, existing rejuvenation approaches consider ageing based only on memory leaks. Future research considering the effects of other ageing-related bugs (e.g., unreleased locks, unterminated threads) will be helpful to properly quantify ageing effects, which is required to properly schedule rejuvenation. Furthermore, the performance of existing tools in the area needs to be enhanced. For instance, studies that compared tools for detecting API-compatibility issues did not show a clear winner [88]. Each tool has its own benefits, and thus future research should focus on hybrid approaches combining existing tools to help mitigate limitations of individual tools and enhance the performance of such tools. Furthermore, more research is required to enable an understanding of how reliability impacts, or is impacted by, other factors such as other quality attributes or the sustainability of applications, which will be important to assess the overall quality and sustainability of apps over time.

**CRediT authorship contribution statement**
**Chathrie Wimalasooriya**: Literature review, Methodology, Data curation and results, Discussion, Writing – original draft. **Sherlock A. Licorish**: Supervision, Literature review, Methodology, Data curation and results, Discussion, Writing – review & editing. **Daniel Alencar da Costa**: Supervision, Literature review, Methodology, Data curation and results, Discussion, Writing – review & editing. **Stephen G. MacDonell**: Supervision, Methodology, Discussion, Writing – review & editing.

**Declaration of competing interest**
This research was funded by AHEAD Operation, a World Bank funded project to accelerate higher education expansion and development in Sri Lanka.

**Chathrie Wimalasooriya** is currently pursuing a Ph.D. degree at University of Otago, New Zealand. She received a Masters in Software Engineering from University of Canterbury, New Zealand and Bachelors in Computing and In- formation Systems from Sabaragamuwa University, Sri Lanka. She has been a lecturer in the Department of Computing and Information Systems and Sabaragamuwa University since 2015. Prior to entering academia, she was in the software development industry as a software engineer from 2013 to 2015. Her research interests are repository mining, empirical software engineering, software maintenance and machine learning applications

**Sherlock A. Licorish** is a Senior Lecturer in the Department of Information Science at University of Otago, New Zealand. He was awarded his Ph.D. by Auckland University of Technology (AUT), and joined University of Otago in 2014. His research portfolio covers agile methodologies, practices and processes, teams and human factors, human computer interaction, research methods and techniques, software code quality, static analysis tools, machine learning applications and software analytics. Sherlock occupies several service roles across the university, nationally and internationally.

**Daniel Alencar da Costa** is a Lecturer (Assistant Professor) in the Department of Information Science at the University of Otago, New Zealand. Daniel obtained his Ph.D. in Computer Science at the Federal University of Rio Grande do Norte (UFRN) in 2017 followed by a Postdoctoral Fellowship at Queen's University, Canada, from 2017 to 2018. His research goal is to advance the body of knowledge of Software Engineering methods and practices through empirical studies, incorporating statistical and machine learning based approaches as well as consulting and documenting the experience of software developers.

**Stephen G. MacDonell** is Professor of Software Engineering at Auckland Uni- versity of Technology and Professor in Information Science at the University of Otago, both in New Zealand. Stephen was awarded BCom(Hons) and MCom degrees from the University of Otago and a Ph.D. from the University of Cambridge. He is a Fellow of IT Professionals NZ, Senior Member of the IEEE and the IEEE Computer Society, and Member of the ACM, and he serves on the Editorial Board of Information and Software Technology. Stephen is also Theme Leader for Data Science & Digital Technologies in New Zealand's National Science Challenge Science for Technological Innovation, Technical Advisor to the Office of the Federation of Ma̅ori Authorities Pou Whakata̅more Hangarau - Chief Advisor Innovation & Research, and Deputy Chair of Software Innovation New Zealand (SINˆZ).